\documentclass[aps,prd,reprint,groupedaddress]{revtex4-2}

\usepackage{physics}
\usepackage{amsmath,amsthm,amssymb}
\usepackage{mathrsfs}
\usepackage{graphicx}
\usepackage{dcolumn}
\usepackage{bm}
\usepackage{siunitx}
\usepackage{caption}
\usepackage{subcaption}
\usepackage[hidelinks]{hyperref}
\usepackage{mathtools}

\begin{document}
\title{Regular Power-Maxwell Black Holes}
	
\author{Yi-Bo Liang}
  \email[]{liangyibo@stu.xjtu.edu.cn}
\author{Hong-Rong Li}
 \email[]{hrli@xjtu.edu.cn}
\affiliation{School of Physics, Xi’an Jiaotong University, Xi’an 710049, China}
	
\date{\today}
\begin{abstract}
We present a new class of regular, spherically symmetric spacetimes in nonlinear electrodynamics that are asymptotically dynamical but not de Sitter, exhibiting power-law Maxwell behavior at infinity. Generalizing to black holes, we derive their existence conditions and construct corresponding Penrose diagrams. Both the weak and dominant energy conditions are shown to be satisfiable. Magnetic solutions are first obtained, with electric counterparts derived via FP duality. Uniqueness conditions for the electric solutions are then established. Although electric duals are absent in square-root Maxwell theory, our auxiliary scalar formulation restores duality and enables a generalized duality transformation. The effective light propagation metric remains regular for particular magnetic configurations (without black holes) but becomes singular for electric cases. Additionally, spacelike photon trajectories are admitted in this spacetime. Finally, the ADM mass is shown to enter the Lagrangian, with the first law and Smarr formula derived, establishing the existence of thermodynamically stable black holes with positive heat capacity.

\end{abstract}
	
\maketitle
	\section{Introduction.}
	\label{int}
	Nonlinear electrodynamics (NED), extending Maxwell's theory beyond its linear framework, emerged in the 1930s with two foundational motivations.
	Born and Infeld pioneered the field by formulating a theory \cite{born1934foundations} that implies the modifications to Maxwell's equations they introduced suffice to resolve the problem of infinite self-energy.
	Much later, a Born-Infeld-like theory turned out to be an important ingredient of String Theory \cite{fradkin1985non,metsaev1987born,seiberg1999string,gibbons2001aspects}
	On the other hand, Heisenberg and Euler developed a distinct NED framework, prompted by Dirac’s theory of the positron (according to which an electromagnetic field can spontaneously create particle-antiparticle pairs, thereby modifying Maxwell’s equations even in vacuum) \cite{heisenberg1936folgerungen,heisenberg2006consequences}.
	Since then, these and other nonlinear electrodynamics models developed over decades have been extensively studied and applied across diverse subfields of theoretical physics, with notable applications in general relativity (GR).
	
	NED-GR solutions with electric or magnetic fields are currently widely discussed \cite{pellicer1969nonlinear,bronnikov1979scalar,bronnikov1976reissner}.
	Beyond Born-Infeld and Heisenberg-Euler theories, one finds diverse nonlinear electrodynamics models including logarithmic, double-logarithmic, exponential, rational, arcsin, power-law, inverse, quasitopological, and other variants \cite{sorokin2022introductory,soleng1995charged,ayon1998regular,ayon2000bardeen,bronnikov2000comment,bronnikov2001regular,burinskii2002new,hassaine2007higher,jing2011holographic,hendi2012asymptotic,gaete2014remarks,kruglov2016nonlinear,kruglov2017black,liu2020quasi,gullu2021double,balakin2021towards,dehghani2022self,fan2016construction}.
	In the context of black holes, the celebrated singularity theorems established by Penrose and Hawking \cite{sw1973large} demonstrate that under certain conditions, singularities are inevitable within classical general relativity. Nevertheless, singularities are widely regarded as nonphysical artifacts of classical gravitational theories, unlikely to exist in nature. Quantum arguments advanced by Sakharov \cite{sakharov1966nachal} and Gliner \cite{gliner1966algebraic} further suggest that spacetime singularities may be avoidable when matter sources feature de Sitter cores at their centers.
	Based on this idea, Bardeen proposed the first static spherically symmetric regular black hole solution \cite{bardeen1968non} without specifying the underlying theory whose field equations it satisfies.
	Only significantly later were the first exact electric and magnetic regular black hole solutions derived and analyzed in specific models \cite{ayon1998regular,ayon2000bardeen,bronnikov2000comment,bronnikov2001regular,burinskii2002new}.
	It was also shown \cite{bronnikov2001regular} that purely magnetic regular configurations can be easily constructed when $L(f)$ (where $f=F_{ab}F^{ab}$) approaches a finite limit as $f\rightarrow\infty$.
	Electric solutions with the same metric are also obtainable, but they exhibit multivaluedness in $L(f)$ \cite{bronnikov2001regular}.
	Subsequently, it was established that the Kiselev geometry \cite{kiselev2003quintessence,lungu2025charged} becomes an exact solution within power-Maxwell electrodynamics \cite{rodrigues2022bardeen,dariescu2022kiselev}.
	Power-Maxwell-Schwarzschild black holes \cite{dariescu2022kiselev,dariescu2024charged} and power-Maxwell-Bardeen black holes \cite{rodrigues2022bardeen} have been investigated.
	Power-Maxwell black holes are particularly noteworthy because, in certain configurations, they can exhibit asymptotic dynamical behavior without being de Sitter.
	However, these power-Maxwell spacetimes contain singularities; consequently, power-Maxwell black holes remain singular regardless of whether the metric includes a Schwarzschild term or a Bardeen term.
	
	In this paper, we present a new class of regular spherically symmetric spacetimes that are asymptotically dynamical but not de Sitter.
	Specifically, these solutions in nonlinear electrodynamics exhibit asymptotic power-Maxwell behavior.
	Magnetic solutions are first constructed and shown to satisfy the weak energy condition; with an additional condition imposed, the dominant energy condition is likewise fulfilled.
	These solutions are then extended to include black holes, while preserving the regularity of the entire spacetime.
	We identify the necessary conditions for the existence of such black holes and present the corresponding, novel Penrose diagrams of these spacetimes.
	By employing the well-established FP duality, we subsequently derive the associated electric solutions and establish uniqueness conditions for their existence.
	However, it is observed that, although magnetic configurations exist in the square-root Maxwell theory, their electric duals do not.
	To resolve this issue, we reformulate the nonlinear theory within the framework of an auxiliary scalar field, thereby enabling the consistent inclusion of dual solutions.
	Within this extended formalism, we derive a generalized duality transformation.
	The effective metric governing light propagation in nonlinear electrodynamics is subsequently examined.
	For magnetic solutions, we establish the conditions under which the effective metric remains regular, whereas for electric solutions, we identify unavoidable singularities.
	The presence of spacelike photon trajectories is uncovered within these spacetimes, introducing fundamental causality ambiguities that warrant further investigation.
	The ADM mass naturally emerges in the Lagrangian formulation, enabling derivation of both the first law of black hole thermodynamics and the Smarr formula.
	This framework establishes the existence of thermodynamically stable black hole solutions characterized by positive heat capacity.
	
	The paper is organized as follows.
	In Sec.~\ref{sub:2a},we construct a family of magnetically charged spacetimes and present complete Penrose diagrams illustrating their evolution—from power-Maxwell geometries to regular power-Maxwell spacetimes, and ultimately to regular power-Maxwell black holes.
	In Sec.~\ref{sub:2b}, electric solutions are derived via FP duality; uniqueness conditions for these solutions are established, alongside a counterexample highlighting the incompleteness of FP duality.
	Section~\ref{sub:2c} introduces an auxiliary field formulation and proposes a novel duality framework that extends beyond FP duality.
	In Secs.~\ref{sub:3a} and~\ref{sub:3b}, we analyze light propagation in magnetic and electric solutions, respectively.
	In Sec.~\ref{sub:4a} and Sec.~\ref{sub:4b}, we derive the first law of thermodynamics and the Smarr formula.
	In Sec.~\ref{sub:4c} we examine the system’s heat capacity and demonstrate the existence of thermodynamically stable black holes with positive heat capacity.
	We conclude in Sec.~\ref{con}.
	In the following, we employ $c=G=1, 4\pi\varepsilon_0=1$ units.

	\section{Regularized Power-Maxwell Black Holes}
	\label{sec:2}
	
	\subsection{The magnetic solutions}
	\label{sub:2a}
	Within the magnetic ansatz framework, the action for power-Maxwell electrodynamics with positive exponent $n> 0$ takes the form
	\begin{equation}
		S=\frac{1}{16\pi}\int_M\epsilon\left[R- L(f) \right]=\frac{1}{16\pi}\int_M\epsilon\left[R- 2\lambda \left(\frac{f}{2\lambda}\right)^n\right],\label{PL}
	\end{equation}
	where $\epsilon$ is the volume element, $\lambda$ is a positive constant with dimensions of inverse length squared, i.e., $[L^{-2}]$.
	Here, $f=F_{ab}F^{ab}$ defines the electromagnetic invariant constructed from the field strength tensor $F$.
	In this configuration, it is evident that for $n=0$, the Lagrangian reduces to $L(f)=2\lambda$, corresponding to a cosmological constant $\lambda$; whereas for $n=1$, $L(f)=f$, which recovers the standard Maxwell Lagrangian..
	Moreover, an explicit dependence of $L$ on the magnetic charge $g$ is excluded, as $g$ should appear as an integration constant, while $L$ is expected to involve only fundamental constants \cite{bronnikov2001regular}.
	Suppose the static, spherically symmetric metric is given by
	\begin{align}
		g=-A(r)\mathrm{d}t^2+\frac{\mathrm{d}r^2}{A(r)}+r^2(\mathrm{d}\theta^2+\sin^2\theta\mathrm{d}\varphi^2).
	\end{align}
	In this case, the field strength tensor takes the form $F=g\sin\theta\mathrm{d}\theta\wedge\mathrm{d}\varphi$, and the electromagnetic invariant satisfies $f=2g^2/r^4$ identically, where $g$ denotes the magnetic charge.
	The solution to Eq.~\eqref{PL} is given by
	\begin{equation}
		A(r)=1-\frac{\lambda^{1-n} g^{2n}}{\gamma}r^{2-4n}-\frac{2M}{r}\label{PLS},\gamma\equiv3-4n
	\end{equation}
	Firstly, we consider the case $M=0$.
	Our main focus is on solutions that are asymptotically dynamical, i.e., those satisfying $A(r)<0$ as $r\rightarrow\infty$.
	Accordingly, we restrict our attention to the parameter ranges $0< n < 1/2$ and $n=1/2$ with $\sqrt{\lambda}g>1$.
	Although solutions with $n<0$ (corresponding to a different Lagrangian, to satisfy weak energy condition, we must have $\mathrm{d}L/\mathrm{d}f\leq0$ \cite{bokulic2021black}), are also asymptotically dynamical, but the corresponding energy density diverges as $r\to +\infty$, and thus such cases are excluded from consideration.
	
\begin{figure}[b]      
		\centering             
		\includegraphics[width=0.5\textwidth]{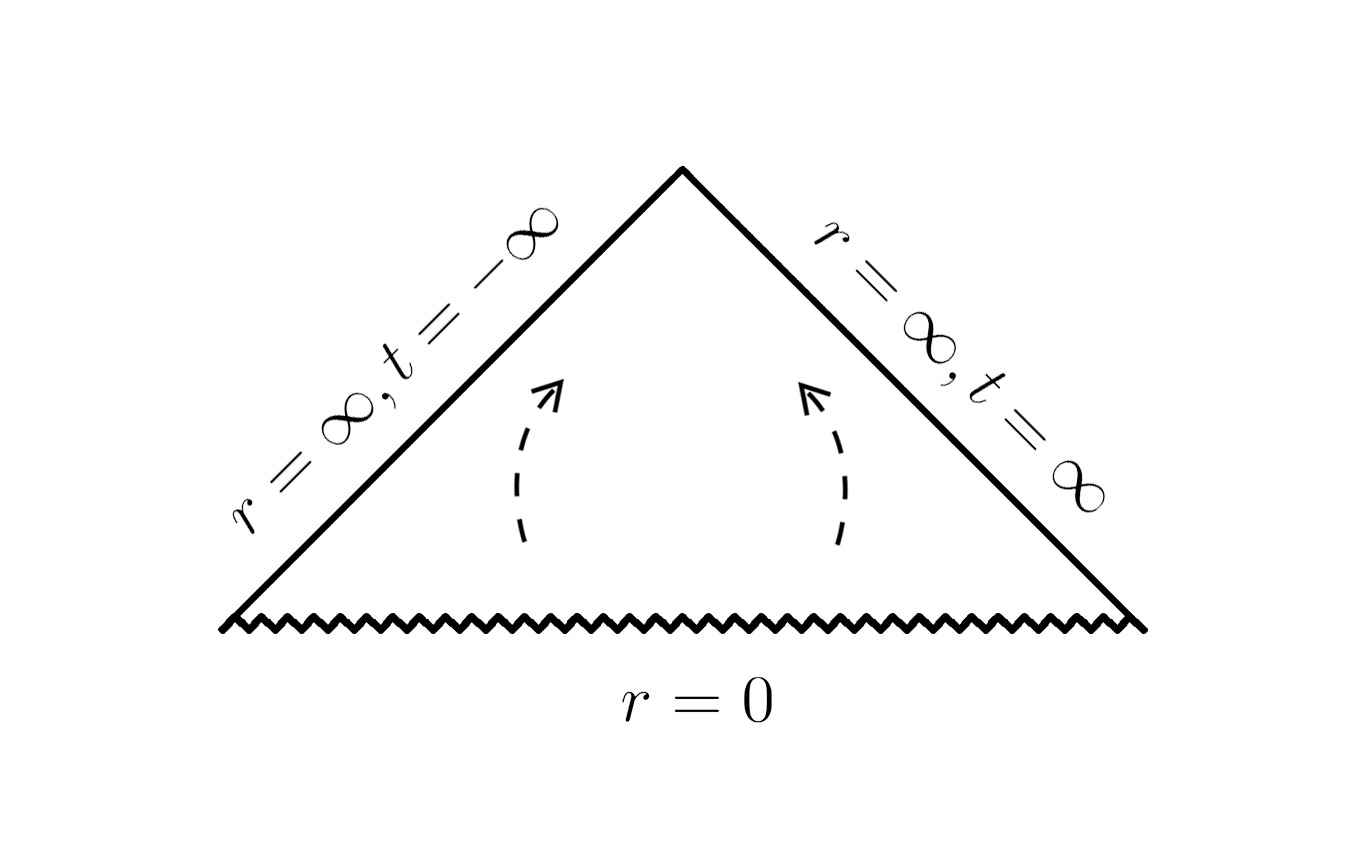}
		\caption{Penrose diagram for $n=1/2$ with $m=0$. The dashed arrows represent the vector field $\partial/\partial r$, while the zigzag line indicates the curvature singularity.}
		\label{fig:1}    
	\end{figure}

\begin{figure}[htbp]      
		\centering             
		\includegraphics[width=0.5\textwidth]{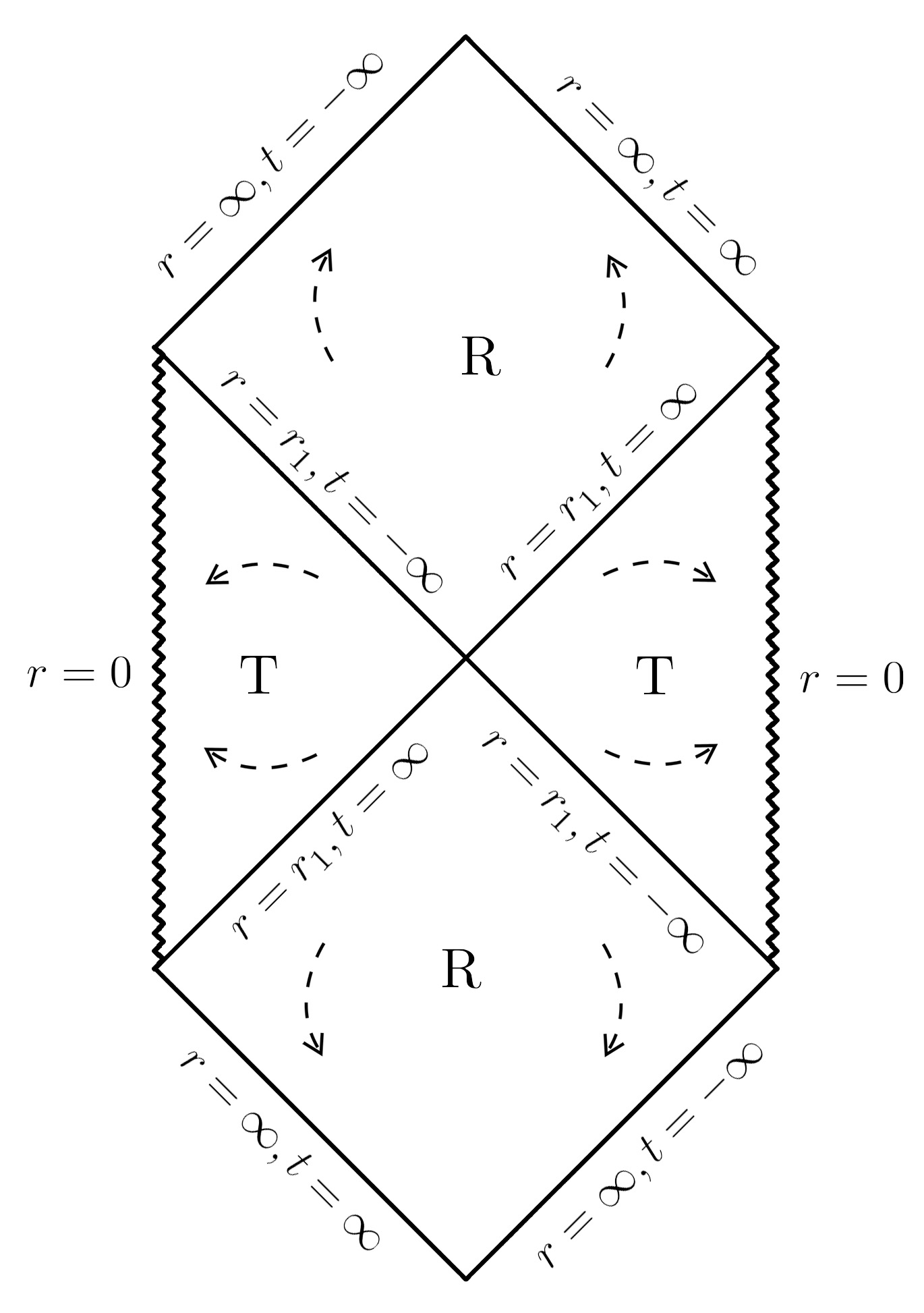}
		\caption{Penrose diagram for $0<n<1/2$ with $m=0$. The dashed arrows represent the vector field $\partial/\partial t$, while the zigzag line indicates the curvature singularity.}
		\label{fig:2}    
	\end{figure}

	In such cases, it should be noted that the spacetimes do not possess topological defects but do exhibit a curvature singularity at $r=0$.
	Under the conformal transformation \cite{penrose1974relativistic} with conformal factor $\Omega=1/r$, the metric function transforms as $A(r)/r^2=1/r^2-\lambda^{1-n} g^{2n}/(\gamma r^{4n})$.
	As $r\rightarrow\infty$, indicating the presence of future and past null infinities.
	The corresponding Penrose diagrams (at constant $\theta$ and $\phi$) are shown in Figs.~\ref{fig:1} and~\ref{fig:2}.
	
	For $n=1/2$ and $\sqrt{\lambda}g>1$, the spacetime depicted in Fig.~\ref{fig:1} is globally hyperbolic and spatially homogeneous.
	The curvature singularity at $r=0$ is gobally naked and resembles a cosmological singularity. 
	Moreover, the spacetime is dynamically evolving and spatially homogeneous, since the vector $\partial/\partial r$ is timelike.
	There is uniform expansion on the two‑sphere, whereas the $t$-direction (which is spacelike) exhibits no expansion.
	In contrast, for $0<\sqrt{\lambda}g<1$, the spacetime becomes non‑dynamical and possesses a timelike singularity at $r=0$.
	If $\sqrt{\lambda}g=1$, the metric is degenerate.
	
	For $0<n <1/2$, the spacetime exhibits a cosmological-like horizon.
	The causal structure, as shown in Fig.~\ref{fig:2}, resembles a rotated version of the Schwarzschild spacetime.
	The central singularity is locally naked.
	In the upper $\mathrm{R}$ region, define $\mathrm{d}\tau\equiv\mathrm{d}r/\sqrt{-A}$, one finds that $\mathrm{d}r/\mathrm{d}\tau\equiv \mathrm{d}a_{\Omega}/\mathrm{d}\tau>0$ and $\mathrm{d}^2a_{\Omega}/\mathrm{d}\tau^2=\lambda^{1-n} g^{2n}(1-2n)r^{1-4n}/\gamma>0$, indicating an accelerating expansion of the two-sphere. 
	Along the $t$ direction, we find $\mathrm{d}\sqrt{-A}/\mathrm{d}\tau\equiv\mathrm{d}a_{t}/\mathrm{d}\tau=\mathrm{d}^2r/\mathrm{d}\tau^2>0$, from which it follows that the acceleration is positive for $0<n<1/4$, vanishes at $n=1/4$, and becomes negative for $1/4<n<1/2$.
	As $r\to +\infty$, we have $a_{\Omega}=r\propto\tau^{1/(2n)}$ and $a_{t}=\sqrt{-A}\propto \tau^{-1+1/(2n)}$.
	Therefore, only when $n=0$, there will be an exponential and uniform expansion as $r\to +\infty$.
	When $0<n\leq 1/2$, the expansion is relatively weak.
	
	To regularize the singularities, it is important to note that the regularity of $L(f)$ alone is insufficient, as argued in \cite{bronnikov2023regular}.
	A more effective approach is to directly assume a regular form for the metric.
	In general, one may consider
	\begin{equation}
	 	A(r)=1-\frac{\lambda b^{4n}}{\gamma}\frac{r^{2+\mu}}{(r^\nu+b^\nu)^{\frac{4n+\mu}{\nu}}},\label{M1}
	\end{equation}
	where $\mu\geq0,\nu>0$ and $b^2=g/\sqrt{\lambda}$, such that the metric remains regular at $r=0$ while asymptotically approaching the power-Maxwell metric as $r\rightarrow \infty$.
	However, to satisfy the weak energy condition, one must impose $\mu=0$.
	The dominant energy condition can also be satisfied.
	For the metric given by Eq.~\eqref{M1} with $\mu=0$, we have $T_{tt}/A(r)-T_{rr}A(r)\equiv\rho-p_r=2\rho\geq 0$ and $T_{\theta\theta}/r^2=T_{\varphi\varphi}/(r^2\sin^2\theta)\equiv p_\theta$.
	Consequently, the only condition required to satisfy the dominant energy condition is $\rho-p_\theta\geq0$, which lead to the condition $0<\nu\leq [6-7n+2\sqrt{3(4n^2-7n+3)}]/n>\gamma$.
	Then, we can derive the Lagrangian of NED \cite{ayon2000bardeen,bronnikov2001regular,fan2016construction} as
	\begin{equation}
		L(f)=\frac{2\lambda}{\gamma}\frac{3[f/(2\lambda)]^{n+\nu/4}+\gamma[f/(2\lambda)]^n}{\{1+[f/(2\lambda)]^{\nu/4}\}^{1+4n/\nu}},\label{LFL}
	\end{equation}
	since for $A(r)=1-2m(r)/r$, the Lagrangian takes the form $L=4m^\prime(r)/r^2$.
	It can be observed that when $n=0$, we have $L(f)=2\lambda$, implying that no regularization occurs in this case.
	The regularized spacetime (at constant $\theta$ and $\phi$) shares the same causal structure as depicted in Fig.~\ref{fig:2} (which is determined by the propagation of light and will be examined in the next section), except for that a point on the line with $r=0$ in Fig.~\ref{fig:2} is now a pole of a 3 sphere \cite{borde1994open,borde1997regular}.
	
	
	Furthermore, it is natural to introducing a Bardeen-type, Hayward-type, or Fan-Wang--type modification \cite{ayon2000bardeen,hayward2006formation,fan2017critical,fan2016construction} into Eq.\eqref{M1}, thereby yielding a regular black hole spacetime.
	The metric satisfying the weak energy condition is given by
	\begin{equation}
		A(r)=1-\frac{\lambda b^{4n}}{\gamma}\frac{r^{2}}{(r^\nu+b^\nu)^{\frac{4n}{\nu}}}-\frac{2\alpha q^3 r^2}{(r^\sigma+q^\sigma)^{\frac{3}{\sigma}}},\label{RPMB}
	\end{equation}
	where $\sigma>0$, $q^2=g/\sqrt{\alpha}$ and $\alpha$ is a positive constant with dimension $[L^{-2}]$.
	In this case, setting $\lambda=0$, yields an ADM mass of $\mathcal{M}=\alpha q^3$.
	In this spacetime, if $0<\nu\leq [6-7n+2\sqrt{3(4n^2-7n+3)}]/n>\gamma$ and $\sigma>1$, both weak and dominant energy condistions will be satisfied.
	The Lagrangian of NED is given by
	\begin{align}
		L(f)=&\frac{2\lambda}{\gamma}\frac{3[f/(2\lambda)]^{n+\nu/4}+\gamma[f/(2\lambda)]^n}{\{1+[f/(2\lambda)]^{\nu/4}\}^{1+4n/\nu}}\notag\\&+\frac{12\alpha[f/(2\alpha)]^{3/4+\sigma/4}}{\left[1+[f/(2\alpha)]^{\sigma/4}\right]^{1+3/\sigma}}.\label{RMBPL}
	\end{align}
	It should be note that whether $\lambda=0$ or $\alpha=0$, the Lagrangian is well-defined.
	One may equivalently cast $A$ in the form
	\begin{equation}
		A(r(\tilde{r}))=1-\frac{\tilde{\lambda} }{\gamma}\frac{\tilde{r}^2}{(\tilde{\lambda}^{\nu/4}\tilde{r}^\nu+1)^{\frac{4n}{\nu}}}-\frac{2\tilde{\alpha}\tilde{r}^2}{(\tilde{\alpha}^{\sigma/4}\tilde{r}^\sigma+1)^{\frac{3}{\sigma}}},\label{RMBPD}
	\end{equation}
	where $\tilde{r}\equiv r/g, \tilde{\lambda}\equiv \lambda g^2,\tilde{\alpha}\equiv \alpha g^2$.
	It is evident that setting $\alpha=\lambda$ simplifies the function $A$ and reduces the number of parameters in the Lagrangian; however, this relation will not be assumed here.
	The equation $A=0$ cannot, in general, be solved in closed analytic form.
	Nevertheless, one can still derive useful necessary conditions governing the existence and number of its real roots.
	
	(i). Noting that $A\rightarrow1$ as $r\rightarrow0$ while $A<0$ as $r\to\infty$, a genuine black hole solution—with inner, event and cosmological‑like horizons—requires $A$ to vanish three times in the interval $(0,+\infty)$, this in turn forces the existence of exactly two extrema for $r>0$.
	Since
	\begin{gather}
		\frac{\mathrm{d}A}{\mathrm{d}\tilde{r}}=-\frac{2\tilde{\lambda}}{\gamma}\tilde{r}R_1(\tilde{r})+2\tilde{\alpha}\tilde{r}R_2(\tilde{r}),\label{DADR}\\
		R_1(\tilde{r})\equiv\frac{(1-2n)\tilde{\lambda}^{\nu/4}\tilde{r}^\nu+1}{(\tilde{\lambda}^{\nu/4}\tilde{r}^\nu+1)^{1+4n/\nu}},\\
		R_2(\tilde{r})\equiv\frac{\tilde{\alpha}^{\sigma/4}\tilde{r}^\sigma-2}{(\tilde{\alpha}^{\sigma/4}\tilde{r}^\sigma+1)^{1+3/\sigma}},
	\end{gather}
	demanding that  $\mathrm{d}A/\mathrm{d}\tilde{r}$ vanish at two distinct radii is equivalent to the condition that the following equation
	\begin{equation}
		R(\tilde{r})\equiv\frac{R_2(\tilde{r})}{R_1(\tilde{r})}=\frac{\tilde{\lambda}}{\gamma\tilde{\alpha}},\label{R12}
	\end{equation}
	has two solutions.
	One then finds that real roots occur only for $\tilde{r}>2^{1/\sigma}\tilde{\alpha}^{-1/4}$.
	As $\tilde{r}\to(2^{1/\sigma}\tilde{\alpha}^{-1/4})^+$, one has $R(\tilde{r})\rightarrow 0^+$.
	In the limit $\tilde{r}\rightarrow+\infty$, $R(\tilde{r})$ behaves as $R(\tilde{r})\propto \tilde{\lambda}^{n}\tilde{\alpha}^{-3/4}\tilde{r}^{4n-3}/(1-2n)\rightarrow 0^+$.
	Consequently, $R(\tilde{r})$ attains its global maximum at some $\tilde{r}=\tilde{r}_M$, and one must therefore require that its maximal value satisfy $R(\tilde{r}_M)>\tilde{\lambda}/(\gamma\tilde{\alpha})$.
	Roughly speaking, since $R_1(\tilde{r})$ is a monotonic increasing positive function, and $R_2(\tilde{r})$ has a positive maximum, if $R_1(\tilde{r})$ changes slowly while $R_2(\tilde{r})$ changes sharply, the location of the global maximum of $R(\tilde{r})$ can be approximated by that of $R_2(\tilde{r})$ approximately.
	It follows from a straightforward calculation that $R_2(\tilde{r})$ attains its global maximum at $\tilde{r}_{M_2}=\tilde{\alpha}^{-1/4}(\sigma+2)^{1/\sigma}$.
	Substitute $\tilde{r}_{M_2}$ into Eq. \eqref{R12} then yields the relation
	\begin{equation}
		\tilde{\lambda}<\frac{\sigma\gamma\tilde{\alpha}[(\tilde{\lambda}/\tilde{\alpha})^{\nu/4}(\sigma+2)^{\nu/\sigma}+1]^{1+4n/\nu}}{(\sigma+3)^{1+3/\sigma}[(\tilde{\lambda}/\tilde{\alpha})^{\nu/4}(1-2n)(\sigma+2)^{\nu/\sigma}+1]}.\label{LM}
	\end{equation}
	We denote by $\lambda_{M_2}$ the value of $\lambda$ for which the above inequality becomes an equality..
	
\begin{figure}[htbp]
		\centering
		
		\begin{subfigure}[b]{0.45\textwidth}
			\includegraphics[width=\textwidth]{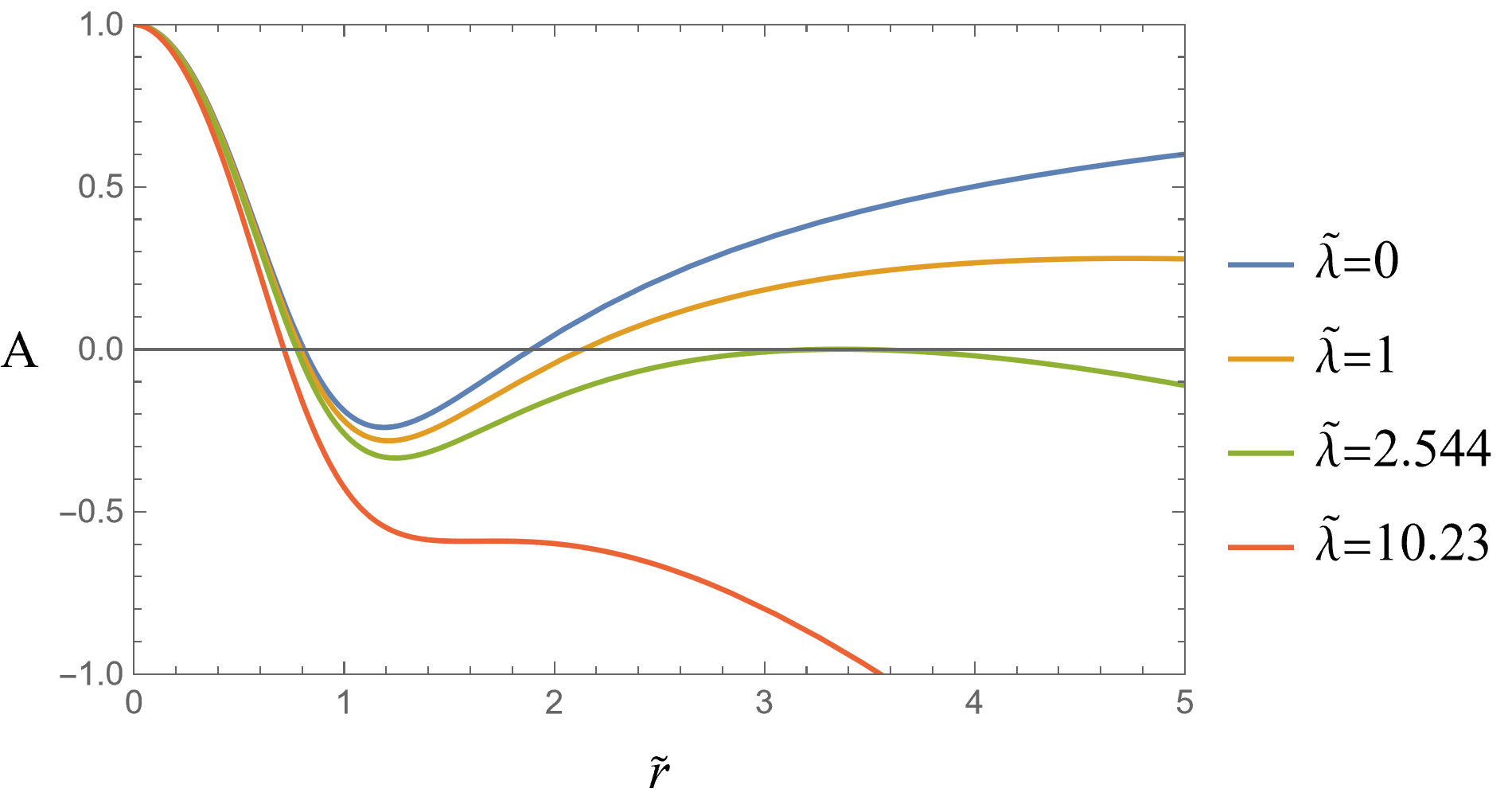}
			\caption{$\nu=1/4,\sigma=4,\tilde{\alpha}=1.$ It can be seen that when $\tilde{\lambda}=0$, $A$ possesses two zeros. In the special case when $\tilde{\lambda}\approx 2.544$, $A$ exhibits exactly two zeros, with one zero occurring at an extremum- this corresponds to an extremal black hole configuration. For $\tilde{\lambda}$ approxinately in $(0,2.544)$, there are three zeros of $A$ since $A<0$ as $r\rightarrow+\infty$. Furthermore, if $\tilde{\lambda}=10.23\approx\tilde{\lambda}_{M_2}$, $A$ tends to have no extrema.}
			\label{fig:z1}
		\end{subfigure}
		
		\vspace{0.2cm}
		
		\begin{subfigure}[b]{0.45\textwidth}
			\includegraphics[width=\textwidth]{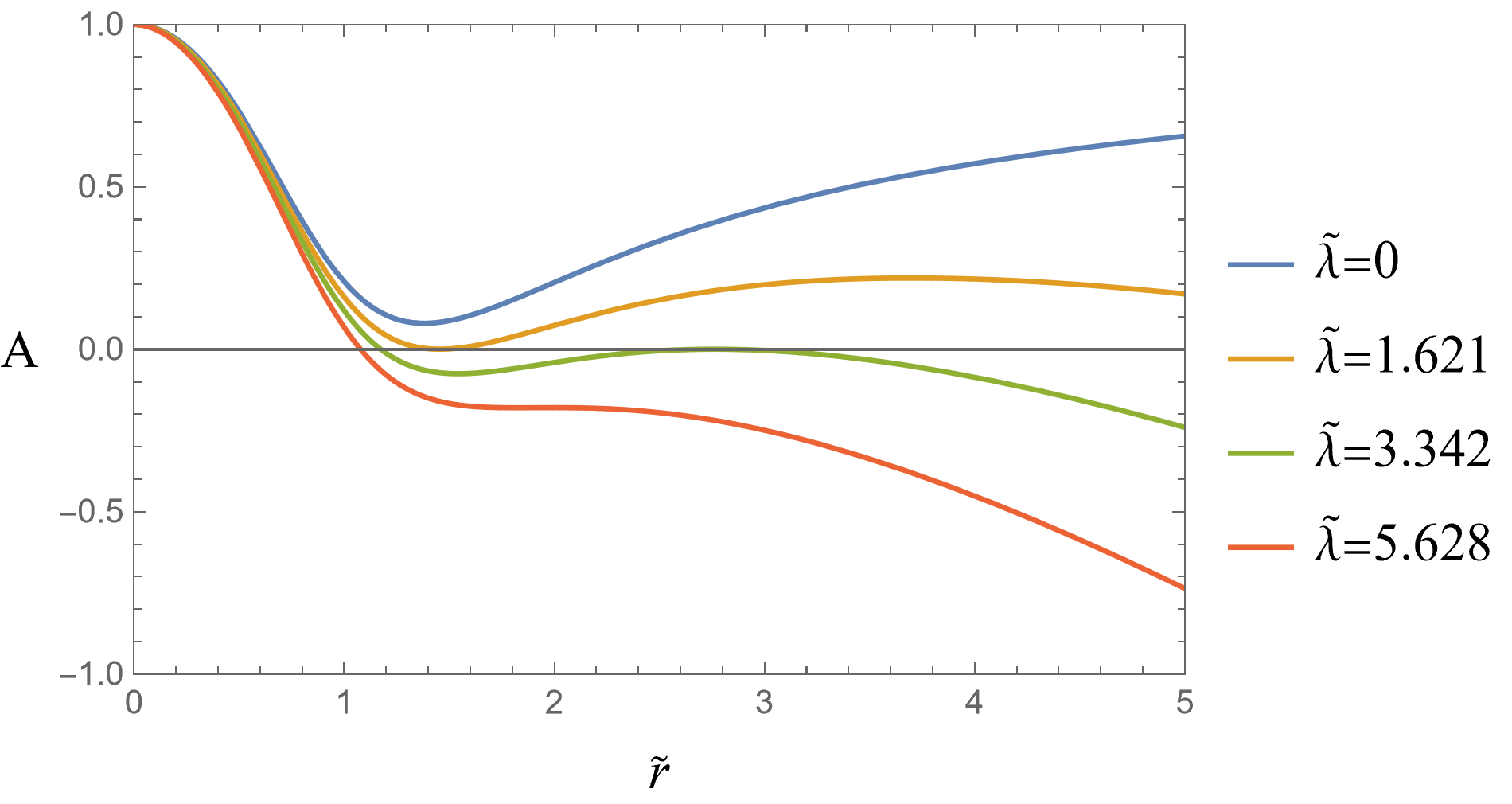}
			\caption{$\nu=1/4,\sigma=4,\tilde{\alpha}=0.55.$ For $\tilde{\lambda}=0$, $A$ exhibits no zeros. At the critical value  $\tilde{\lambda}\approx 1.621$ and $\tilde{\lambda}\approx 3.342$, the solution corresponds to two distinct classes of extremal black holes. The metric function $A$ develops three zeros for $\tilde{\lambda}$ approximately in $(1.621,3.342)$. Furthermore, if $\tilde{\lambda}=5.628\approx\tilde{\lambda}_{M_2}$, $A$ tends to have no extrema.}
			\label{fig:z2}
		\end{subfigure}
		
		\caption{$A$ associated to the solution Eq.~\eqref{RMBPD} with $n=1/4$.}
		\label{fig:z}
	\end{figure}

	(ii). The derivative $\mathrm{d}A/\mathrm{d}\tilde{r}$ must be negative in the asymptotic limit $\tilde{r}\rightarrow+\infty$.
	For the specific case where $0<n<1/2$, we obtain
	\begin{equation}
		\frac{\mathrm{d}A}{\mathrm{d}\tilde{r}}\sim-\frac{2\tilde{\lambda}^{1-n}(1-2n)}{\gamma \tilde{r}^{4n-1}}+\frac{2\tilde{\alpha}^{1/4}}{\tilde{r}^2}.
	\end{equation}
	Thus, the asymptotic behavior of $\mathrm{d}A/\mathrm{d}\tilde{r}$ is dominated by a negative leading term, with no additional constraints required.
	Figure~\ref{fig:z} displays the characteristic behavior of the metric function $A$ for $n=1/4$, comparing two physically distinct scenarios: (1). when $\tilde{\lambda}=0$ with a black hole present, and (2). when $\tilde{\lambda}=0$ without a black hole.
	Let $r_2$ denote the second zero of $A$.
	Since the $\tilde{\lambda}$-dependent term in $A$ is negative, the value of $r_2$ increases monotonically with increasing $\tilde{\lambda}$ (when $\nu$ is determined).
	
	If $n=1/2$, as $\tilde{r}\to+\infty$, we have
	\begin{align}
	\frac{\mathrm{d}A}{\mathrm{d}\tilde{r}}\sim&-\frac{2\tilde{\lambda}^{1/2-\nu/4}}{\tilde{r}^{1+\nu}}\left(1-\frac{1+2/\nu}{\tilde{\lambda}^{\nu/4}\tilde{r}^\nu}\right)\notag\\&+\frac{2\tilde{\alpha}^{1/4}}{\tilde{r}^2}\left(1-\frac{3+3/\sigma}{\tilde{\alpha}^{\sigma/4}\tilde{r}^\sigma}\right).
	\end{align}

\begin{figure}[!htb]
		\centering
		
		\begin{subfigure}[b]{0.45\textwidth}
			\includegraphics[width=\textwidth]{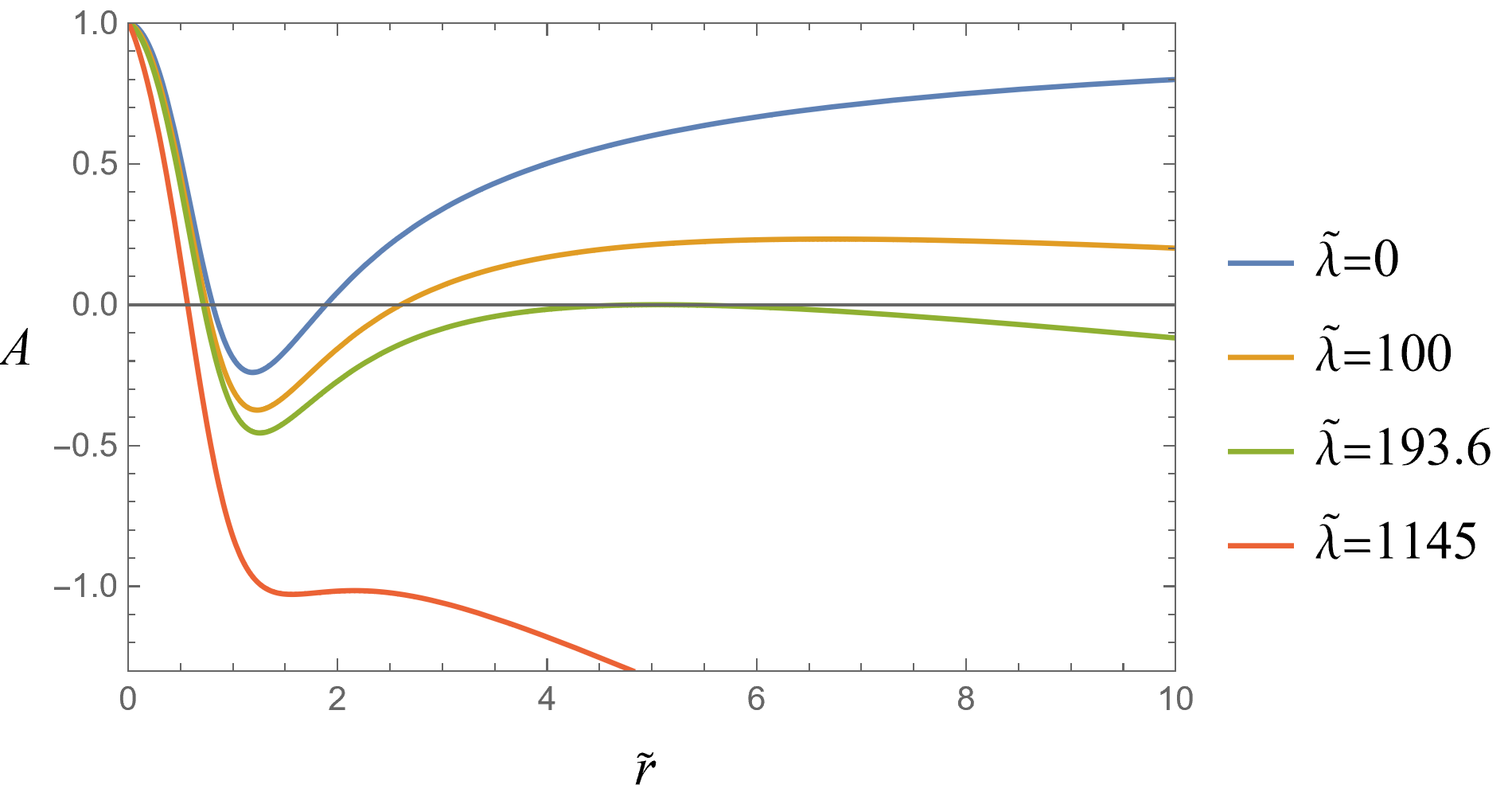}
			\caption{$\nu=1/4,\sigma=4,\tilde{\alpha}=1.$ It can be seen that when $\tilde{\lambda}=0$, there are two zeros of $A$. For $\tilde{\lambda}\approx 193.6$, the solution corresponds to an extremal black hole. When $\tilde{\lambda}$ lies approximately in the interval $(1,193.6)$, $A$ possesses three distinct zeros. Furthermore, if $\tilde{\lambda}=1145\approx\tilde{\lambda}_{M_2}$, $A$ tends to have no extrema.}
			\label{fig:l1}
		\end{subfigure}
		
		\vspace{0.2cm}
		
		\begin{subfigure}[b]{0.45\textwidth}
			\includegraphics[width=\textwidth]{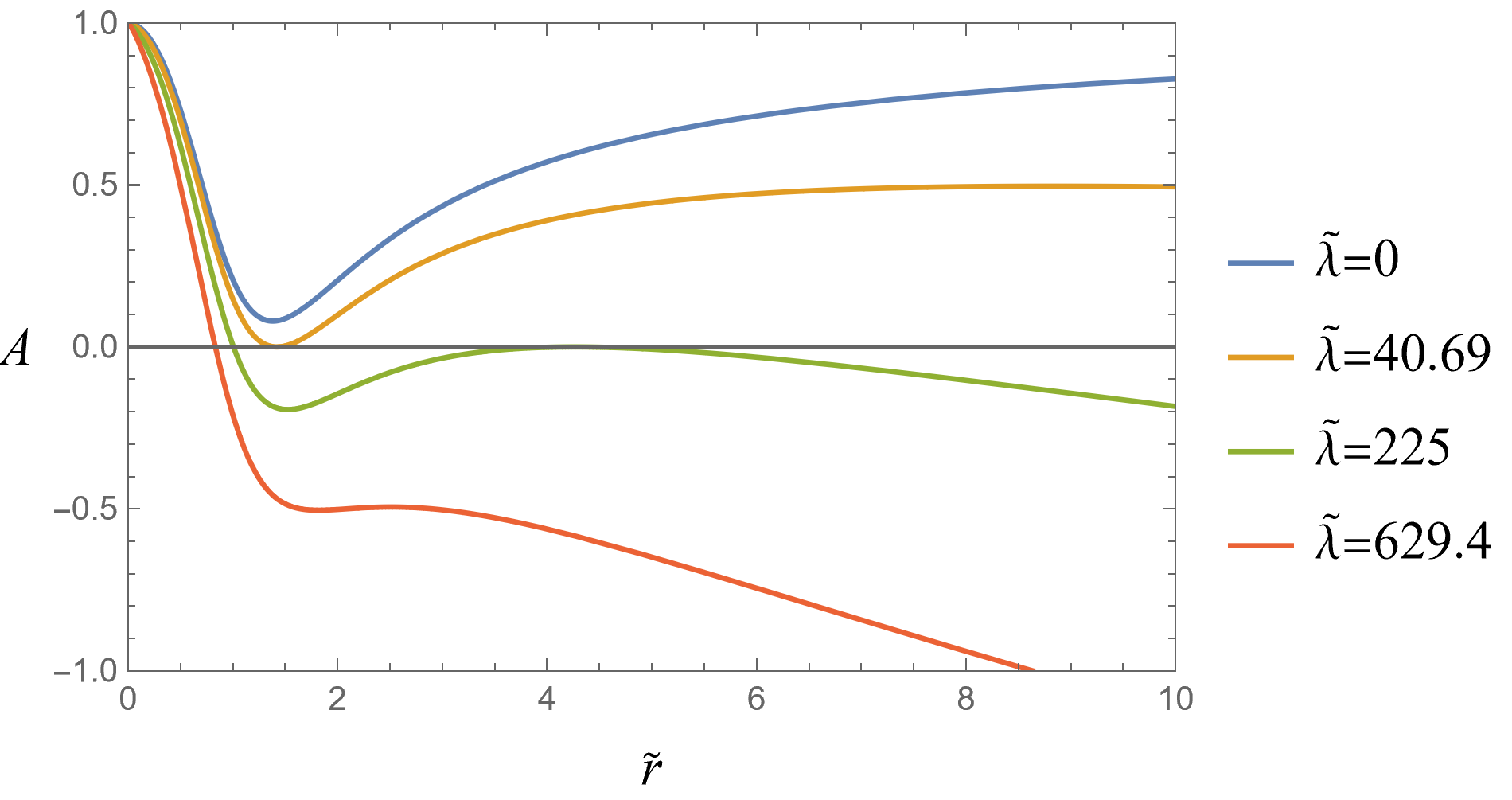}
			\caption{$\nu=1/4,\sigma=4,\tilde{\alpha}=0.55.$ When $\tilde{\lambda}=0$, there are no zeros of $A$. If $\tilde{\lambda}\approx 40.69$ or $\tilde{\lambda}\approx 225$, there are different extremum black holes. When $\tilde{\lambda}$ lies approximately in the interval $(40.69,225)$, $A$ possesses three distinct zeros. Furthermore, if $\tilde{\lambda}=629.4\approx\tilde{\lambda}_{M_2}$, $A$ tends to have no extermums.}
			\label{fig:l12}
		\end{subfigure}
		
		\vspace{0.2cm}
		
		\begin{subfigure}[b]{0.45\textwidth}
			\includegraphics[width=\textwidth]{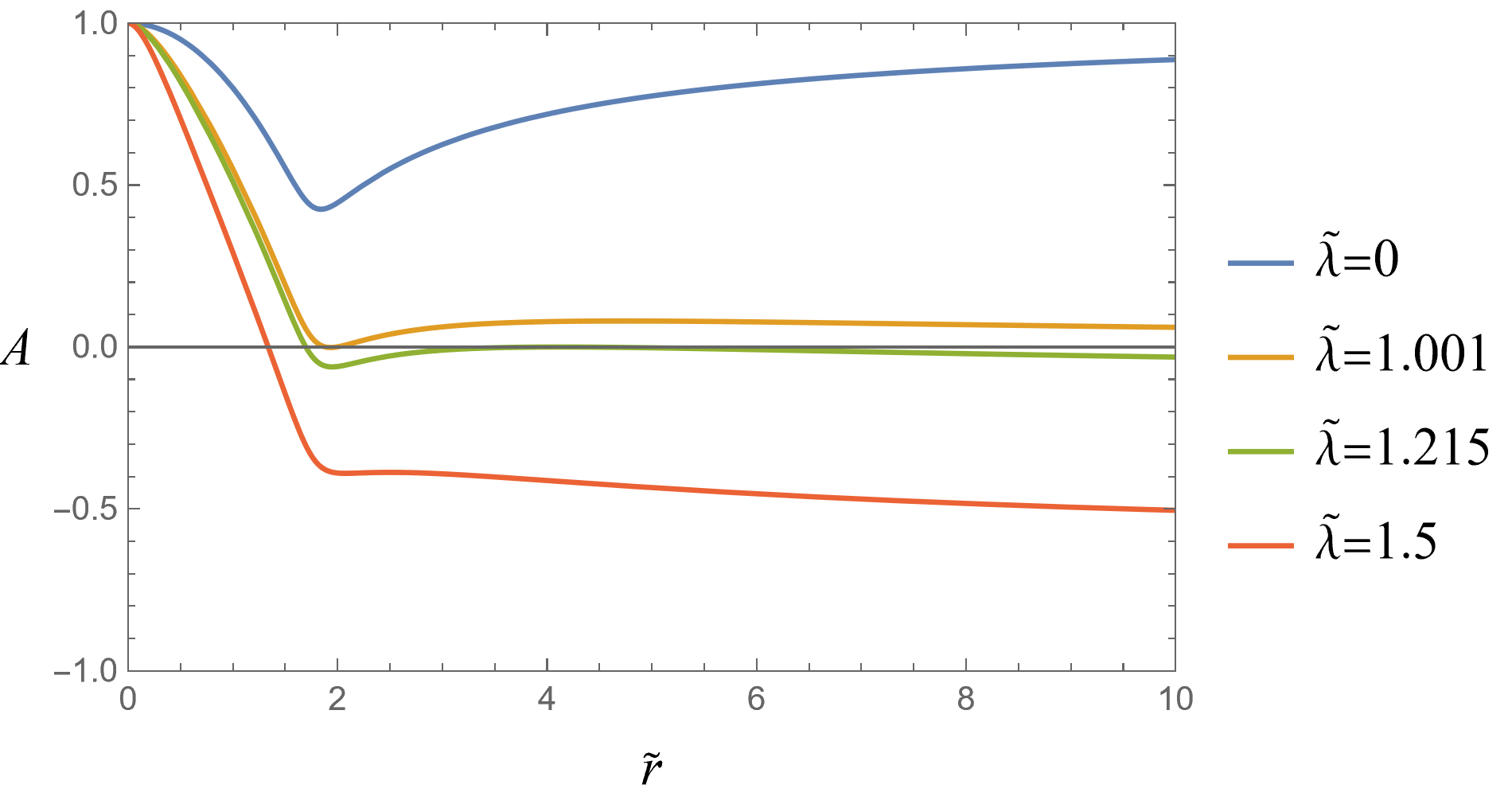}
			\caption{$\nu=1,\sigma=20,\tilde{\alpha}=0.1.$ When $\tilde{\lambda}=0$, there are no zeros of $A$. If $\lambda\approx 1.001$ or $\tilde{\lambda}\approx 1.215$, there are different extremum black holes. When $\tilde{\lambda}$ lies approximately in the interval $(1.001,1.215)$, $A$ possesses three distinct zeros. Furthermore, if $\tilde{\lambda}=1.5\approx\tilde{\lambda}_{M_2}$, $A$ tends to have no extermums.}
			\label{fig:l2}
		\end{subfigure}
		
		\caption{When $n=1/2$, the behaviour of $A$ is shown.}
		\label{fig:l}
	\end{figure}

	(ii-1). Clearly, if $0<\nu<1$, then $\mathrm{d}A/\mathrm{d}\tilde{r}$ is negative, while for $\nu>1$, it becomes positive.
	The behavior of $A$ is illustrated in Figs.~\ref{fig:l1} and~\ref{fig:l12} for the case $0<\nu<1$.
	
	(ii-2). For the case $\nu=1$ with $\tilde{\lambda}<\tilde{\alpha}$, we have $\mathrm{d}A/\mathrm{d}\tilde{r}>0$, which violates the requirement and must therefore be excluded from consideration.
	
	(ii-3). For the case $\nu=1$ with $\tilde{\lambda}=\tilde{\alpha}$, the condition $0<\sigma\leq1$ must be imposed to guarantee $\mathrm{d}A/\mathrm{d}\tilde{r}<0$.
	In this case, we have
	\begin{equation}
		\frac{\mathrm{d}A}{\mathrm{d}r}=-2\tilde{\lambda}\tilde{r}\left[\frac{1}{(\tilde{\lambda}^{1/4}\tilde{r}+1)^3}-\frac{\tilde{\lambda}^{\sigma/4}\tilde{r}^\sigma-2}{(\tilde{\lambda}^{\sigma/4}\tilde{r}^\sigma+1)^{1+3/\sigma}}\right].
	\end{equation}
	It remains to see the sign in the expression inside the square brackets.
	For $\tilde{\lambda}^{\sigma/4}\tilde{r}^\sigma\leq2$, this term is positive; thus, it remains only to determine the sign of the term in the regime $\tilde{\lambda}^{\sigma/4}\tilde{r}^\sigma>2$.
	Letting $x\equiv \tilde{\lambda}^{\sigma/4}\tilde{r}^\sigma$, we are led to determine the sign of $N(x)$, with
	\begin{equation}
		N(x)\equiv(x+1)^{1+3/\sigma}-(x-2)(x^{1/\sigma}+1)^3,x>2.
	\end{equation}
	Define $W(x)$ as
	\begin{equation}
		W(x) \equiv \ln[(x+1)^{1+3/\sigma}]-\ln[(x-2)(x^{1/\sigma}+1)^3],
	\end{equation}
	then we have $N(x)=[\mathrm{e}^{W(x)}-1](x-2)(x^\sigma+1)^3$, hence, $W(x)>0\iff N(x)>0$.
	Since $W\rightarrow+\infty$ as $x\rightarrow2^+$, and for $x\rightarrow +\infty$, $W(x)\propto(3+3/\sigma)/x-3 x^{-1/\sigma}$, it follows that $W(x)\to 0^+$ as $x\to +\infty$, given that $0<\sigma\leq 1$.
	Hence, the sign of $\mathrm{d}W/\mathrm{d}x$ should be determined.
	It follows naturally that
	\begin{equation}
		\frac{\mathrm{d}W(x)}{\mathrm{d}x}=\frac{1+3/\sigma}{x+1}-\frac{1}{x-2}-\frac{3x^{1/\sigma-1}}{\sigma(x^{1/\sigma}+1)}.
	\end{equation}
	After some simplification, $\mathrm{d}W/\mathrm{d}x$ is found to share the same sign as $J(x)$, where
	\begin{equation}
		J(x)\equiv[2-(1+\sigma)x]x^{1/\sigma-1}+(x-2-\sigma).
	\end{equation}
	Since $x>2$, it follows that $2-(1+\sigma)x<-\sigma x$.
	To analyze the sign of $V(x)\equiv(x-2-\sigma)-\sigma x^{1/\sigma}>J(x)$, we compute its derivative and obtain $V^\prime(x)=1-x^{1/\sigma-1}$.
	Given that $0<\sigma\leq1$, it follows $1/\sigma-1\geq 0$, and hence $x^{1/\sigma-1}\geq 1$ for all $x>2$. Thus, $V^\prime(x)\leq 0$, indicating that $J$ is monotonically decreasing or constant for $x>2$. 
	Since $V(x)<0$ as $x\rightarrow2^+$, we conclude that $V(x)<0$ for all $x>2$, hence $J(x)<0$ for all $x>2$.
	As a result, $W(x)$ is a monotonically decreasing function of $x$, hence it is always positive, which implies $N(x)>0$ and consequently $A^\prime(r)<0$.
	Therefore, $A(r)$ has at most one zero, and the spacetime does not contain a black hole.
	
	\begin{figure*}[htbp]      
	\centering             
	\includegraphics[width=0.8\textwidth]{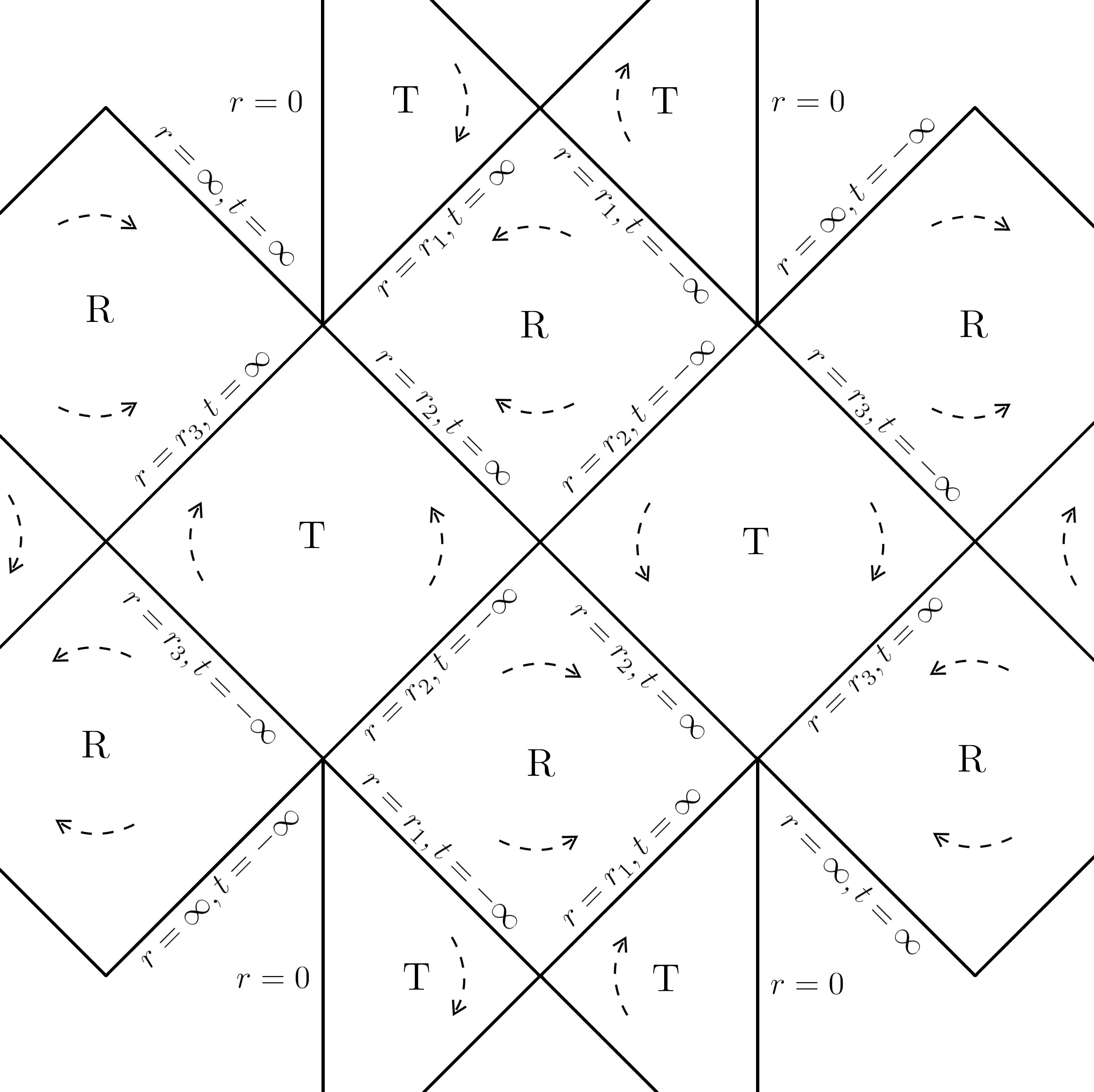}
	\caption{Penrose diagram of the regularized Black hole sourced by a power‑law Maxwell field, in the case where $A(r)$ admits three distinct positive roots. The dashed arrows indicate the timelike Killing vector field $\partial/\partial t$, and the two points at $r=0$ separated by the same inner horizon located at $r=r_1$ are the antipodal points of a 3 sphere. The Penrose diagram is unbounded in both the horizontal and vertical directions.}
	\label{fig:6}    
\end{figure*}
	
	(ii-4). If $\nu=1$ and $\tilde{\lambda}>\tilde{\alpha}$, then $\mathrm{d}A/\mathrm{d}r<0$ holds throughout, and black hole solutions do exist in this case, as illustrated in Fig.~\ref{fig:l2}.
	In this parameter regime, we have not found configurations exhibiting behavior analogous to those illustrated in Figs.~\ref{fig:l1} and~\ref{fig:z1}.
	
	Comparing with Figs~\ref{fig:z} and~\ref{fig:l}, it is observed that, under the same parameters ($\nu,\sigma$ and $\tilde{\alpha}$), the case $n=1/2$ permits a larger range of $\tilde{\lambda}$.
	The Penrose diagram, corresponding to the case with three distinct zeros of $A$ (indicating the presence of a black hole), is shown in Fig.~\ref{fig:6}.
	
	\subsection{The electric solutions}
	\label{sub:2b}
	By applying the FP duality transformation \cite{bronnikov2001regular,moreno2003stability},
	\begin{equation}
		\{F,f,L(f)\}\longleftrightarrow\{\star P,-p,-H(p)\},\label{FP}
	\end{equation}
	one readily obtains the corresponding purely electric configurations.
	Here, we define $P\equiv L_f F, p\equiv P_{ab}P^{ab}, H(p)\equiv2fL_f-L$, where $L_f\equiv\mathrm{d}L/\mathrm{d}f$, and $\star$ denotes the Hodge dual.
	Then, the following relations are valid: $L=2pH_p-H, L_f H_p=1, f=pH_p^2, p=fL_f^2$.
	By employing the FP duality map, one can derive an electrically charged configuration from magnetic $L(f)$ via either of the following, fully equivalent, routes: (i) perform a Legendre transform of $L(f)$ and obtain magnetic $H(p)$, then directly apply the duality to $H(p)$ to obtain electric $L(f)$; or (ii) first apply the duality to $L(f)$ to obtain electric $H(p)$, and thereafter reconstruct electric $L(f)$.
	The choice of procedure depends on whether magnetic $H(p)$ is given explicitly as a function of $p$; if it is not, one should employ the latter method to obtain the electric theory in the $P$ framework.
	However, FP duality does not always guarantee the existence of a corresponding electric solution.
	In the case of a magnetic configuration with $H(p)=0$, it can be solved to yield
	\begin{equation}
		L(f)=a\sqrt{f},
	\end{equation}
	where $a$ is a positive constant with dimension $[L^{-1}]$.
	Therefore, the first approach to obtain the electric $L(f)$ is not feasible.
	Furthermore, following the second approach, yields electric $H(p)=-a\sqrt{-p}$, which implies $L=2pH_p-H$=0.
	Therefore, for $n=1/2$, no dual electric theory exists; in other words, there is no dual solution corresponding to Eq.~\eqref{PLS}.
	Conversely, there is no dual magnetic theory corresponding to the electric theory with $L(f)=-a\sqrt{-f}$.
	It should be noted, however, that no solution exists for the electric theory $L(f)=-a\sqrt{-f}$ under the static, spherically symmetric metric.
	
	Nevertheless, we will now show that the dual electric solution for $n=1/2$ can indeed be derived through an alternative methodological approach.
	It is evident that an arbitrary nonlinear electrodynamics Lagrangian $L(f)$ can be constructed from a Maxwell field coupled to an auxiliary scalar field $\phi$, as described by the Lagrangian $\mathcal{L}(f,\phi)$:
	\begin{gather}
		\mathcal{L}(f,\phi)=h(\phi)f+j(\phi),\label{LFP}\\
		h(\phi)\equiv L_\phi, j(\phi)\equiv L-L_\phi \phi, L\equiv L(f(\phi)), \phi\equiv f.\label{PS}
	\end{gather}
	One can verify that Eq.~\eqref{PS} satisfies the equation of motion for the scalar field $\phi$, as derived from the Lagrangian $\mathcal{L}(f,\phi)$ defined in Eq.~\eqref{LFP}.
	An important special case arises when $j(\phi)=0$, which immediately yields the linear solution $L(\phi)=a\phi$, where $a$ is a dimensionless constant.
	This solution corresponds to Maxwell's theory with an additional scaling coefficient $a$.
	We now demonstrate that the solution space of the $\mathcal{L}(f,\phi)$ system is strictly larger than that of the $L(f)$ system.
	In particular, the $\mathcal{L}(f,\phi)$ framework admits dual electric solutions corresponding to the metric in Eq.~\eqref{PLS} even for $n=1/2$, a case where the conventional $L(f)$ formulation fails to yield dual solutions.
	Under FP duality, the electric sector Lagrangian is
	\begin{equation}
		L(f)=2\lambda (1-2n)\left(-\frac{f}{2\lambda n^2}\right)^{\frac{n}{2n-1}},
	\end{equation}
	which is dual to the magnetic Lagrangian.
	For the parameter range $0<n<1/2$, we observe that $n/(2n-1)$ satisfies $-\infty<n/(2n-1)<0$.
	This behavior explicitly demonstrates the nonexistence of dual solutions at the critical value $n=1/2$.
	Applying the field equations given in Eq.~\eqref{PS} yields the relation
	\begin{gather}
		h(\phi)=\frac{1}{n}\left(-\frac{\phi}{2\lambda n^2}\right)^{\frac{1-n}{2n-1}},\\ j(\phi)=2\lambda(1-n)\left(-\frac{\phi}{2\lambda n^2}\right)^{\frac{n}{2n-1}}.
	\end{gather}
	Define
	\begin{gather}
		h(\phi(\psi))=\frac{1}{n}\psi^{1-n},\label{PMH}\\ j(\phi(\psi))=2\lambda(1-n)\psi^n.\label{PMG}
	\end{gather}
	Hence, the action is well-defined when $n=1/2$, and the solution is 
	\begin{equation}
		A(r)=1-\sqrt{\lambda}g, F=-\frac{\sqrt{\lambda}}{2}\mathrm{d}t\wedge\mathrm{d}r,\psi=\frac{g^2}{\lambda r^4}.\label{TFP}
	\end{equation}
	In this configuration, the electromagnetic invariant evaluates to $f=-\lambda/2$.
	A potential concern arises regarding the $\lambda$ appearing in the denominator of $\psi$, as its vanishing could ostensibly lead to singular behavior.
	However, this apprehension is resolved through careful consideration of the coupled dynamics: the equation of motion for $\psi$ ensures that the total energy-momentum tensor remains well-defined even in the limit $\lambda\to 0$.
	This regularity stems from the fundamental constraint that these functions cannot be treated in isolation due to their interdependence through the field equations.
	Within this solution framework, the parameter $g$ represents the electric charge:
	\begin{equation}
		\frac{1}{8\pi}\int_s \epsilon_{abcd}\left(h(\phi(\psi)) F^{cd}\right)=g. 
	\end{equation}
	While, when $g=0$, the total energy-momentum tensor vanishes, although $F$ and $f$ do not.
	This situation resembles the theory with $L=F_{ab}\star F^{ab}$, where the only nonzero component of $F$ is $F_{tr}$; consequently, the energy-momentum tensor vanishes while $F$ and $f$ remain nonzero.
	Although one could redefine $\lambda\to\lambda^{3/2}g$ to make both $F$ and $f$ vanish, as previously noted, the Lagrangian $L$ may not contain integration constant; therefore, such a redefinition is not adopted.
	It should be kept in mind that this solution cannot be mapped onto a conventional nonlinear electrodynamics theory.
	A detailed examination of the $\mathcal{L}(f,\phi)$ system will be presented in the final subsection of this section.
	
	Attention is now turned to the analysis of the class of regular electric solutions.
	
	(i). If $\alpha=0$, the dual $H$ dual corresponding to Eq.~\eqref{RMBPL} is given by
	\begin{equation}
		H=-\frac{2\lambda}{\gamma}\frac{3[-p/(2\lambda)]^{n+\nu/4}+\gamma [-p/(2\lambda)]^n}{\{1+[-p/(2\lambda)]^{\nu/4}\}^{1+4n/\nu}}.\label{RH}
	\end{equation}
	Then, the quantity $f=p H^2_p$ requires careful examination, as the function $f(p)$ may not be monotonic, and the $P$ framework is equivalent to the $F$ framework only if $f(p)$ is a monotonic  function\cite{bronnikov2023regular}.
	Furthermore, additional issues may arise when $f(p)$ is not monotonic \cite{bronnikov2023regular}, which will be discussed in the following section.
	In this case, we have
	\begin{equation}
		f=-\frac{2\lambda n^2[-p/(2\lambda)]^{2n-1} \{\gamma+(3+\nu)[-p/(2\lambda)]^{\nu/4}\}^2}{\gamma^2\{1+[-p/(2\lambda)]^{\nu/4}\}^{4+8n/\nu}}.
	\end{equation}
	It can be seen that in the limit $p\to 0$ ($r\to +\infty$) with $n=1/2$, it is observed that $f\to -\lambda/2$, consistent with the result presented in Eq.~\eqref{TFP}.
	Letting $u\equiv [-p/(2\lambda)]^{\nu/4}$, and define $w(u)\equiv u^{4(2n-1)/\nu}(1+u)^{-(4+8n/\nu)}[\gamma+(3+\nu)u]^2$.
	To examine the monotonicity of $f(p)$, it is sufficient to analyze the behavior of $w(u)$.
	
	Define $K(u)\equiv\mathrm{d}\ln w/\mathrm{d}\ln u$, if $K(u)\geq 0$ or $K(u)\leq 0$ throughout the domain, then $w$ is monotonic.
	It follows naturally that
	\begin{gather}
		K=\frac{4(2n-1)}{\nu}+\frac{uN(u)}{(1+u)[\gamma+(3+\nu) u]},\\
		N(u)\equiv-(2+\frac{8n}{\nu})[(\gamma-\nu)+(3+\nu) u].
	\end{gather}
	It is observed that in the limit $u\to+\infty$, $K\rightarrow-2(1+2/\nu)<0$; therefore, it is necessary to impose the condition $K(u)\leq 0$ for all $u$.
	For $0<n\leq 1/2$, the condition $N(u)\leq0$ guarantees that $K(u)\leq0$.
	This yields the sufficient condition $\gamma-\nu\geq0$.
	Consequently, if $0<\nu\leq\gamma$, the function $f(p)$ is monotonically decreasing in $p$, and the $P$ framework is equivalent to the $F$ framework.
	Under this condition,$L(f)$ remains single-valued.
	It should be noted that $F=-g/(L_f r^2)\mathrm{d}t\wedge\mathrm{d}r$.

	(ii). If $\lambda=0$ and $\alpha\neq0$, the corresponding dual solution is given by
	\begin{equation}
		H=-\frac{12\alpha[-p/(2\alpha)]^{(3+\sigma)/4}}{\{1+[-p/(2\alpha)]^{\sigma/4}\}^{1+3/\sigma}},\label{36}
	\end{equation}
	where
	\begin{equation}
			f=-\frac{9\alpha (3+\sigma)^2[-p/(2\alpha)]^{(1+\sigma)/2}}{2\{1+[-p/(2\alpha)]^{\sigma/4}\}^{4+6/\sigma}}.
	\end{equation}
	It can be seen that the equation $\mathrm{d}f/\mathrm{d}p\equiv f_p=H_p(H_p+2pH_{pp})=0$ has exactly one root
	\begin{equation}
		p=-2\alpha\left(\frac{1+\sigma}{2+\sigma}\right)^{4/\sigma}.\label{q0}
	\end{equation}
	As $p\to0^+$ and $p\to+\infty$, we have $f\rightarrow0^-$; hence, $f$ cannot be a monotonic function of $q$, and $L(f)$ suffers branching at $q_0$.
	
\begin{figure}[t]
		\centering
		
		\begin{subfigure}[b]{0.45\textwidth}
			\includegraphics[width=\textwidth]{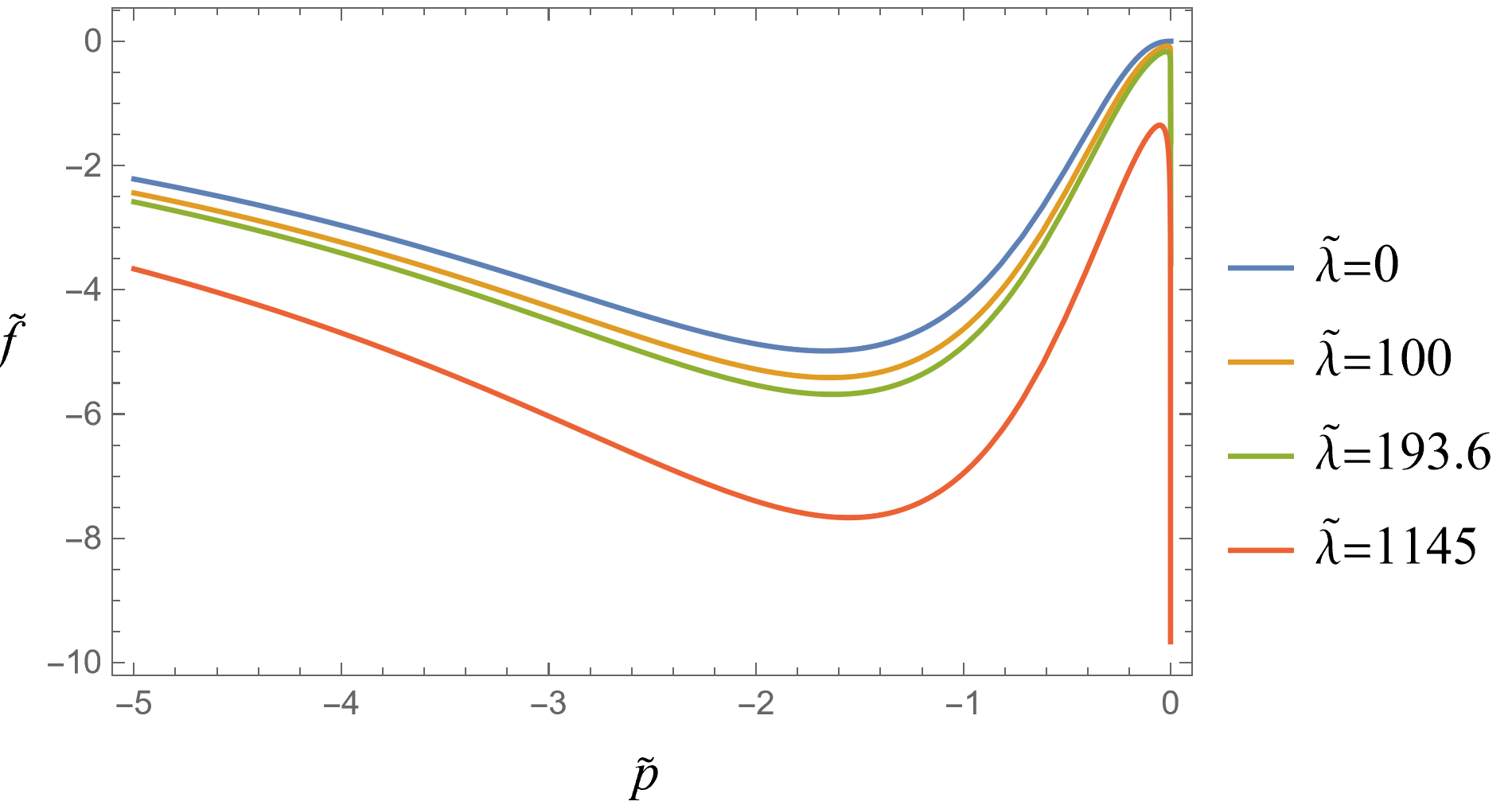}
			\caption{$\nu=1/4,\sigma=4,\tilde{\alpha}=1$.}
			\label{fig:c1}
		\end{subfigure}
		
		\vspace{0.2cm}
		
		\begin{subfigure}[b]{0.45\textwidth}
			\includegraphics[width=\textwidth]{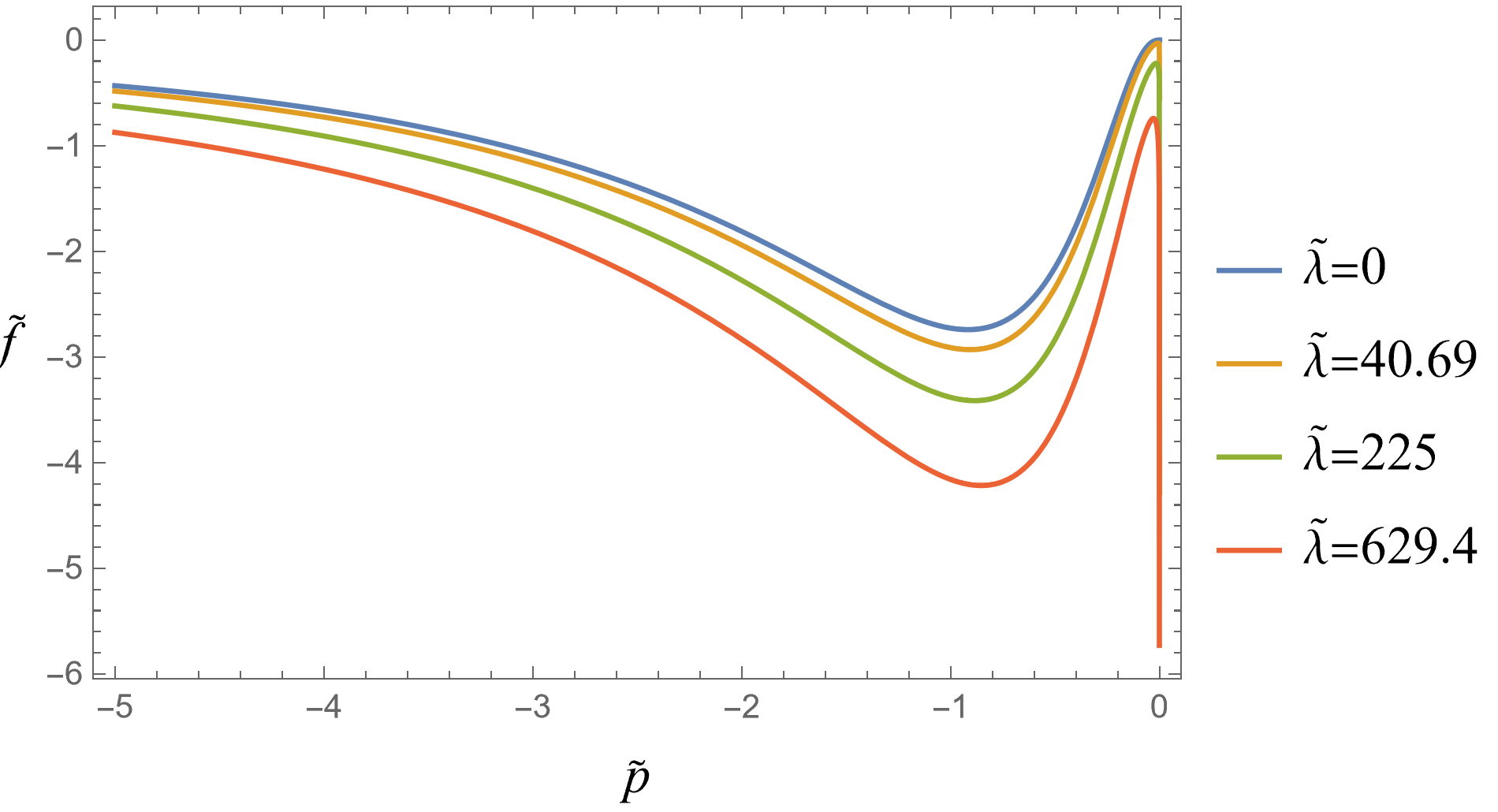}
			\caption{$\nu=1/4,\sigma=4,\tilde{\alpha}=0.55$.}
			\label{fig:c2}
		\end{subfigure}
	
		\vspace{0.2cm}
		
		\begin{subfigure}[b]{0.46\textwidth}
			\includegraphics[width=\textwidth]{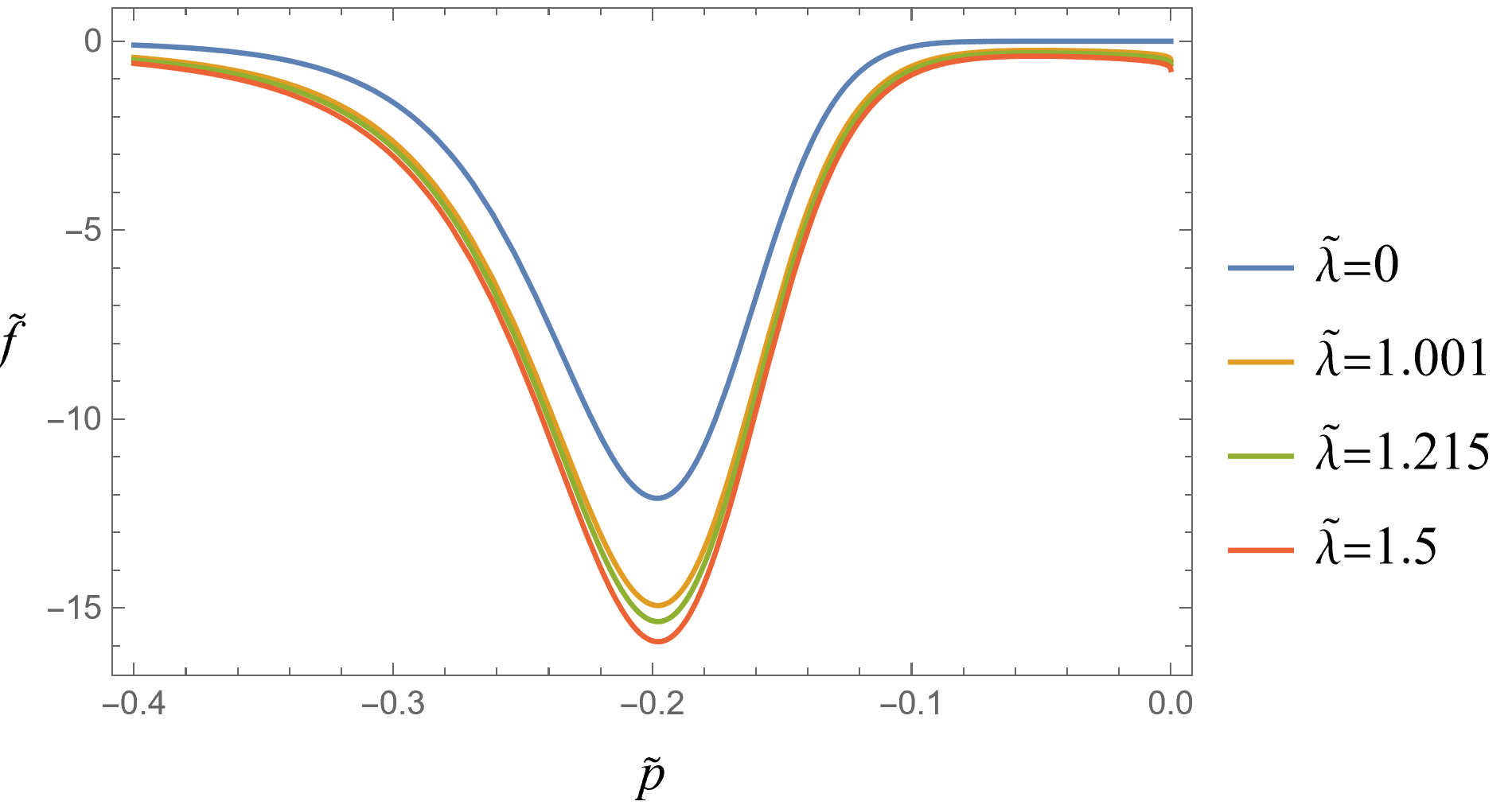}
			\caption{$\nu=1/4,\sigma=20,\tilde{\alpha}=0.1$.}
			\label{fig:c3}
		\end{subfigure}
		
		\caption{The behaviour of $\tilde{f}$ is shown for the case $n=1/2$.}
		\label{fig:c}
	\end{figure}

	(iii). Finally, the monotonicity of $f(p)$ is examined in the case where both $\lambda\neq0$ and $m\neq0$.
	In this case, $H$ is the superposition of Eqs.~\ref{RH} and~\ref{36},
	which renders $f$ too complex for analytical examination.
	The behavior of $g^2f\equiv\tilde{f}$ (rendered dimensionless) for $n=1/2$ is illustrated in Fig.~\ref{fig:c}.
	It can be observed that $\tilde{f}$ fails to be a monotonic function of $g^2 p\equiv\tilde{p}$ in the solution presented.
	For $n=1/4$ with parameters specified in Fig.~\ref{fig:z}, $\tilde{f}$ is also non-monotonic and the corresponding diagram is omitted here.
	
	\subsection{The duality in the $\mathcal{L}(f,\phi)$ formulation.}
	\label{sub:2c}
	
	There is one more point that deserves emphasis: the duality within the $\mathcal{L}(f,\phi)$ formulation.
	Under the assumption of a static, spherically symmetric metric, for magnetic solutions in the form $\mathcal{L}(f,\phi)=h(\phi)f+j(\phi)$, it follows that
	\begin{equation}
		f=-\frac{j_\phi}{h_\phi}=\frac{2g^2}{r^4},F=g\sin\theta\mathrm{d}\theta\wedge\mathrm{d}\varphi.
	\end{equation}
	Hence, one immediately obtains
	\begin{gather}
		T_{tt}=-A^2(r)T_{rr}=\frac{A(r)}{2}\left(-h\frac{j_\phi}{h_\phi}+j\right),\label{T1M}\\
		T_{\theta\theta}=\frac{T_{\varphi\varphi}}{\sin^2\theta}=\frac{r^2}{2}\left(-h\frac{j_\phi}{h_\phi}-j\right)\label{T2M}.
	\end{gather}
	For electric solutions in the form $\mathfrak{L}(f,\psi)=k(\psi)f+l(\psi)$, it follows that
	\begin{equation}
		f=-\frac{l_\psi}{k_\psi}=-\frac{2g^2}{k^2 r^4},
		F=-\frac{g}{kr^2}\mathrm{d}t\wedge\mathrm{d}r.
	\end{equation}
	Then
	\begin{gather}
		T_{tt}=-A^2(r)T_{rr}=\frac{A(r)}{2}\left(k\frac{l_\psi}{k_\psi}+l\right),\label{T1E}\\
		T_{\theta\theta}=\frac{T_{\varphi\varphi}}{\sin^2\theta}=\frac{r^2}{2}\left(k\frac{l_\psi}{k_\psi}-l\right).\label{T2E}
	\end{gather}
	Comparing Eqs.~\eqref{T1M},~\eqref{T2M} with~\eqref{T1E} and~\eqref{T2E}, one can observe that they coincide up to the substitution
	\begin{equation}
	\{F,\phi,j(\phi),\frac{1}{h(\phi)}\}\longleftrightarrow\{\star (kF),\psi,l(\psi),k(\psi)\}.\label{PDU}
	\end{equation}
	Under this duality, the electric solution in the $\mathfrak{L}$ form can be obtained directly.
	Moreover, the monotonicity of $f(\phi)=-j_\phi h^2/h_\phi$ coincides with that of $f(p)$.
	
	Clearly, the reguarized theory given by Eq.~\eqref{RH} is not well-defined in the limit $q\to 0$ when $n=1/2$, since both $H$ and $L$ vanish and then $F$ is not well-defined.
	Hence, the duality in the $\mathcal{L}(f,\phi)$ formulation will be employed to derive a $\mathfrak{L}(f,\phi)$ form from the Lagrangian $L(f)$ in Eq.~\eqref{RMBPL} with $\alpha=0$.
	It follows directly that
	\begin{gather}
		h(\phi)=\frac{n \phi^{n-1} [\gamma+(3+\nu)\phi^{\nu/4}]}{\gamma (1+\phi^{\nu/4})^{2+4n/\nu}},\label{48}\\
		j(\phi)=\frac{ [4n^2+3(1+\phi^{\nu/4})^2-n(7+\phi^{\nu/4}(7+\nu))]}{\gamma (2\lambda \phi^n)^{-1}(1+\phi^{\nu/4})^{2+4n/\nu}}\label{49}.
	\end{gather}
	In the limit $\phi\to0$, one has $h\to n\phi^{n-1}$ and $j\to 2\lambda (1-n)b^n$.
	Under duality, this corresponds to $k\to \phi^{1-n}/n$ and $l\to 2\lambda (1-n)b^n$, which align with Eqs.~\eqref{PMH} and~\eqref{PMG}.
	Hence, the regularized theory in the $\mathfrak{L}$ form requires no expansion and is directly compatible with the solution presented in Eq.~\eqref{TFP}.
	
	\section{Light propagation.}
	\label{sec:3}
	
	As we have seen, the same regular metric of the form given in Eq.\ref{RPMB} can be derived from two distinct sources—electric and magnetic—each governed by a NED theory.
	It is therefore natural to expect that the properties of the electromagnetic field will differ between these two cases.
	According to \cite{novello2000geometrical,novello2000singularities,bronnikov2023regular}, photons governed by NED propagate along null geodesics of the effective metric
	\begin{equation}
		h^{ab}=L_f g^{ab}-4L_{ff}F^a_c F^{cb}.
	\end{equation}
	
	\subsection{The magnetic solutions}
	\label{sub:3a}
	For the metric given by Eq.\ref{RPMB} with a magnetic source, it follows that
	\begin{equation}
		h =-\frac{1}{L_f}\left[A(r)\mathrm{d}t^2-\frac{\mathrm{d}r^2}{A(r)}\right]+\frac{r^2}{\Phi}(\mathrm{d}\theta^2+\sin^2\theta\mathrm{d}\varphi^2),\label{51}
	\end{equation}
	where $\Phi=L_f+2f L_{ff}$.
	
	Initially, the case $\alpha=0$ is considered.
	It is necessary to examine whether singularities arise in the effective metric when $L_f=0$ or $+\infty$, or when $\Phi=0$ or $-\infty$.
	Eq.~\eqref{RMBPL} implies that
	\begin{gather}
		L_f =\frac{n[f/(2\lambda)]^{n-1}\{\gamma+(3+\nu)[f/(2\lambda)]^{\nu/4}\}}{\gamma (1+[f/(2\lambda)]^{\nu/4})^{2+4n/\nu}},\label{Lf}
	\end{gather}
	Since $L_f\geq 0$, it can vanish only at $r=0$, whereupon $\Phi$ also vanishes.
	As $r\to 0$, $L_f\propto r^{4+\nu}$ and $\Phi\propto r^{4+\nu}$; consequently, the Ricci scalar $R$, the Ricci tensor squared $R_{ab}R^{ab}$, and the Kretschmann scalar $\mathcal{K}=R_{abcd}R^{abcd}$ all vanish at $r=0$.
	If $\Phi=0$ at some $r>0$, these points correspond to singularities in the effective metric.
	According to the FP duality, $\Phi$ corresponds to electric quantity $H_p+2pH_{pp}=f_p/H_p$.
	Since electric $H_p$ (equal to magnetic $L_f$) is positive for $r>0$, the zeros of $\Phi$ coincide with the zeros of electric $f_p$.
	As discussed in Sec.~\ref{sub:2b}, the condition $0<\nu\leq \gamma$ ensures that the electric $f(p)$ monotonic decreasing.
	Consequently, electric $f_{p}<0$ for all $r>0$, and hence $\Phi<0$.
	Note that $\Phi<0$ throughout the entire spacetime is a consequence of the choice that $\Phi\neq0$ for $r>0$.
	Moreover, $L_f$ and $\Phi$ can diverge to $+\infty$ and $-\infty$, respectively, only in certain configurations as $r\to +\infty$.
	In the following, without loss of generality, it is assumed that $A>0$.
	
	As noted in \cite{bronnikov2023regular, novello2000singularities}, if an emitter at rest a point $X$ sends a photon with frequency $\omega_X$, it comes to a receiver at rest at point $Y$ with frequency $\omega_Y$ related to $\omega_X$ by
	\begin{equation}
		\frac{\omega_Y}{\omega_X}=\left[\frac{\sqrt{-g_{tt}}}{-h_{tt}}\right]_Y \left[\frac{\sqrt{-g_{tt}}}{-h_{tt}}\right]^{-1}_X=\left[\frac{L_f}{\sqrt{A}}\right]_Y \left[\frac{L_f}{\sqrt{A}}\right]^{-1}_X.\label{FXY}
	\end{equation}
	The relation \eqref{FXY} follows from the Killing equation $\tilde{\nabla}_a(h_{bc}\xi^c)+\tilde{\nabla}_b (h_{ac}\xi^c)=0$ and the geodesic equation $K^a\tilde{\nabla}_aK^b=0$, where $\tilde{\nabla}$ is the covariant derivative operator compatible with the effective metric $h$. $\xi=\partial/\partial t$ is the timelike Killing vector, and $K$ is the photon's four-wavevector.
	If $X$ is a regular point with $r\neq0$ while $Y$ is the regular center ($r=0$), then $L_f|_Y=0$, this means an infinite redshift for photons.
	Hence, the singularity at $\Phi=0$ does not affect the photon frequency.
	The null geodesic equation ($\theta=\pi/2$) in the effective metric is given by
	\begin{gather}
	E^2-\frac{1}{L_f^2}\left(\frac{\mathrm{d} r}{\mathrm{d}\tau}\right)^2-A(r)\Phi\frac{L^2}{r^2}=0,\label{ENG}\\
	E\equiv \frac{A(r)}{L_f}\frac{\mathrm{d}t}{\mathrm{d}\tau}, L\equiv \frac{r^2}{\Phi}\frac{\mathrm{d}\varphi}{\mathrm{d}\tau},
	\end{gather}
	where $\tau$ is the affine parameter, $E$ and $L$ denote the conserved energy and augular momentum of the photon, respectively.
	Clearly, if at a point $r_0$, $L_f\neq0$ and $\Phi=0$, then $\mathrm{d}r/\mathrm{d}\tau$ remains finite while $\mathrm{d}\varphi/\mathrm{d}\tau\to 0$, resulting in the freezing of the $\varphi$ coordinate.
	Therefore, the $\Phi=0$ singularity seems to be unnoticed by the photons.
	Thus, for magnetic solutions, the choice $0<\nu\leq \gamma$ does not appear to be necessary.
	
	It can be seen that radially moving photons propagate at velocity $c$, since the two-dimensional metric of the $(t,r)$ subspace in the effective metric is conformal to that of the metric in Eq.\eqref{RPMB}.
	Hence, the causal structure of the effective spacetime coincides with the Penrose diagram for paths with constant $\theta$ and $\varphi$.
	However, the situation differs for nonradial photon trajectories.
	
	When $\Phi$ becomes negative, the effective metric ceases to be Lorentzian, causing significant alterations to the trajectories of nonradial photon paths.
	In regions where $\Phi<0$, the coordinate $\theta$ and $\varphi$ becomes timelike, allowing photons to propagate within the ($r,\theta$) or ($r,\varphi$) subspaces.
	For example, within the ($r,\varphi$) subspace, the null geodesic equation is given by
	\begin{equation}
		\frac{1}{L_f^2}\left(\frac{\mathrm{d} r}{\mathrm{d}\tau}\right)^2=-A(r)\Phi\frac{L^2}{r^2}.
	\end{equation}
	If at a point $r_0$, $L_f\neq0$ and $\Phi=0$, then $\mathrm{d}r/\mathrm{d}\tau\to0$ and $\mathrm{d}\varphi/\mathrm{d}\tau\to 0$, resulting in the photon being frozen at that location.
	Consequently, their trajectories are spacelike in the original spacetime and correspond to zero energy from the viewpoint of an observer within that spacetime.
	
	Then, in the case $\lambda=0$ with $\alpha\neq0$, we have
	\begin{equation}
		L_f =\frac{3(3+\sigma)[f/(2\alpha)]^{(\sigma-1)/4}}{2\{1+[f/(2\alpha)]^{\sigma/4}\}^{2+3/\sigma}}.
	\end{equation}
	Clearly, $L_f$ and $\Phi$ vanishes at $r=0$, which is not a curvature singularity.
	$L_f$ and $\Phi$ diverge to $+\infty$ only as $r\to+\infty$ when $0<\sigma<1$.
	While, $\Phi$ also vanishes at $q_0=((1+\sigma)/(2+\sigma))^{4/\nu}$; the latter corresponds exactly to the value obtained in Eq.~\eqref{q0} due to FP duality, and it induces a singularity characterized by $h_{\theta\theta}\to+\infty$ or $h_{\theta\theta}\to-\infty$.
	However, this singularity appears to be imperceptible to photons.
	It is noteworthy that $\Phi<0$ in the vicinity of the regular center.
	
	Furthermore, the case where $\lambda\neq0,\alpha\neq0$, and a black hole exists is examined.
	Since $L_f$ and $\Phi$ are linear superpositions of counterparts in the $\alpha=0$ and $\lambda=0$ cases, $L_f>0$ for $r>0$.
	According to FP duality, $\Phi$ shares the same zeros as electric $f_p$ for $r>0$.
	This behavior can, for example, be observed in Fig.~\ref{fig:c}.
	
	The causality issues, along with the behavior of spacelike photons, warrants further investigation.
	
	\subsection{The electric solutions}
	\label{sub:3b}
 	
 	For the metric in Eq.\eqref{RPMB} with an electric source, we have
 	\begin{equation}
 		h =-\frac{1}{\Phi}\left[A(r)\mathrm{d}t^2-\frac{\mathrm{d}r^2}{A(r)}\right]+\frac{r^2}{L_f}(\mathrm{d}\theta^2+\sin^2\theta\mathrm{d}\varphi^2),\label{59}
 	\end{equation}
 	where $\Phi=L_f+2fL_{ff}=H_p/f_p,L_f=1/H_p$.
 	
 	First, the case with $\alpha=0$ is considered.
 	Under the FP duality, $H_p$ corresponds exactly to Eq.\eqref{Lf}; hence, $H_p\geq0$, with equality holding only at $r=0$.
 	As examined in Sec.~\ref{sub:2b}, if $0<\nu\leq \gamma$, then $f(p)$ is a monotonically decreasing function, and $f_p$ vanishes only at $r=0$.
	Therefore, there is no $r>0$ at which $\Phi\to0$ or $\Phi\to-\infty$.
	Hence, only the behavior of the effective metric near $r=0$, where $\Phi\to-\infty$, requires examination.
	This limit corresponds to a curvature singularity in the effective metric.
	Then, for a radially photon travelling from point $X$ to point $Y$, we have
	\begin{equation}
		\frac{\omega_Y}{\omega_X}=\left[\frac{\Phi}{\sqrt{A}}\right]_Y \left[\frac{\Phi}{\sqrt{A}}\right]^{-1}_X.
	\end{equation}
	If $X$ is a regular point (with $\Phi<0$) while $Y$ is located at $r=0$, then any photon arriving there experiences infinite blueshift, gaining an unlimited energy, which thus implies an instability of the entire configuration.
	Since $\Phi<0$ and $L_f>0$ for $r>0$, the effective metric remains Lorentzian, and from the perspective of photons, there is an interchange of the roles of $t$ and $r$.
	Hence, photons may likewise be confined to propagate solely within the ($r,\theta$) or ($r,\varphi$) subspaces.
	
	When $\alpha\neq0$ and a black hole is present, as illustrated in Eq.~\eqref{q0} and Fig.~\ref{fig:c}, $f(p)$ is not monotonic, and $f_p=0$ at the branching points where $H_p$ remains finite; consequently, $\Phi\rightarrow-\infty$ or $\Phi\rightarrow+\infty$ at these points.
	This leads to a curvature singularity in the effective metric and results in an infinite blueshift of photons.
	Moreover, due to the singularity between the region with $\Phi<0$ and the region with $\Phi>0$, it prevents the reception of a photon with negative frequency if the emitted photon possesses positive frequency.
	
	\section{THERMODYNAMICS.}
	\label{sec:4}
	This section primarily focuses on the first law, Smarr formula, and heat capacity of the spacetime.
	For discussions on the zeroth law(s), equation of state, Helmholtz free energy, and related topics, see \cite{gulin2017generalizations,bokulic2021black,rodrigues2022bardeen,singh2020thermodynamics} and the references therein.
	
	\subsection{The first law.}
	\label{sub:4a}
	
	First, it should be note that an integral constant term ($2M/r$) in $A(r)$ has been omitted, corresponding to the choice $M=0$.
	In \cite{fan2016construction}, the parameter $M$ was taken as nonzero to recover the correct first law.
	Conversely, in \cite{ma2014corrected,singh2020thermodynamics,rodrigues2022bardeen}, $M=0$ was assumed, and an additional free parameter $m$ was introduced in the nonlinear electrodynamics Lagrangian to correct the first law.
	In both cases above, $M$ and $m$ are treated as free parameters; while, if $M\neq0$, the spacetime cannot remain regular.
	In this study, $M$ is set to zero, and to ensure that the NED Lagrangian does not contain the integration constant $g$, the ADM mass (for $\lambda=0$) is obtained as $\mathcal{M}=\alpha q^3$, which is not a free parameter.
	The discussion now proceeds to the main content.
	The temperature $T$ of a black hole can be determined from the surface gravity $\kappa$ as prescribed in \cite{wald2001thermodynamics}, with
	\begin{equation}
		T=\frac{\kappa}{2\pi},\kappa^2=-\frac{1}{2}(\nabla^a\xi^b)\nabla_a\xi_b\bigg|_{r=r_2},
	\end{equation}
	where $\xi=\partial/\partial t$ and $r_2$ denotes the event horizon radius, as illustrated in Fig.~\ref{fig:6}.
	Since $\xi$ is the Killing vector, the surface gravity is given by $\kappa=|A^\prime(r)/2|_{r=r_2}$.
	
	Note that when there are three zeros of $A(r)$, the derivative $A^\prime(r)$ evaluated at $r=r_2$ must be positive, and thus we have
	\begin{equation}
		T=\frac{r_2}{2\pi}\left\{\frac{\mathcal{M}(r_2^\sigma-2q^\sigma)}{(r_2^\sigma+q^\sigma)^{1+3/\sigma}}-\frac{\lambda b^{4n}[(1-2n)r_2^\nu+b^\nu]}{\gamma(r_2^\nu+b^\nu)^{1+4n/\nu}}\right\}.\label{TH}
	\end{equation}
	Since the $\lambda$-dependent term in Eq.~\eqref{TH} is strictly negative, the black hole temperature is consequently always lower than that of the solution with $\lambda=0$.
	When $\lambda=0$, the temperature $T$ vanishes at the critical radius $r_2=2^{1/\sigma}g$, corresponding to an extremal black hole configuration.

	In the original parameters configuration, there are three independent parameters: $\lambda,\alpha$ and $g$.
	Additionally, the quantities $b\equiv g^{1/2}\lambda^{-1/4},q\equiv g^{1/2}\alpha^{-1/4}$ and $\mathcal{M}\equiv\alpha q^3= \alpha^{1/4}g^{3/2}$ constitute another set of three linearly independent parameters.
	Then, $\mathcal{M}$ can be determined from the condition $A(r_2)=0$ (noting that $\lambda=\mathcal{M}q b^{-4}$, since the ADM mass was not introduced as a free parameter), which yields
	\begin{equation}
		\mathcal{M}=\frac{1}{r_2^2}\frac{\gamma(r_2^\sigma+q^\sigma)^{3/\sigma}(r^\nu+b^\nu)^{4n/nu}}{b^{4n-4}q (r_2^\sigma+q^\sigma)^{3/\sigma} +2\gamma (r_2^\nu+b^\nu)^{4n/\nu}}.\label{MSbq}
	\end{equation}
	The area law establishes that the black hole entropy is given by $S=A/4=\pi r^2_2$, where $A$ denotes the event horizon area.
	Subsequently, one obtains
	\begin{align}
		\frac{\partial \mathcal{M}}{\partial S}=&\frac{\gamma(r_2^\sigma+q^\sigma)^{-1+3/\sigma}(r_2^\nu+b^\nu)^{-1+4n/\nu}}{\pi r_2^4[b^{4n-4}q(r_2^\sigma+q^\sigma)^{3/\sigma}+2\gamma(r_2^\nu+b^\nu)^{4n/\nu} ]^2}\notag\\&\big\{
		-b^{4n-4}q[(1-2n)r_2^\nu+b^\nu](r^\sigma+q^\sigma)^{1+3/\sigma}\notag\\&+ \gamma(r_2^\sigma-2q^\sigma)(r_2^\nu+b^\nu)^{1+4n/\nu}\big\}.
	\end{align}
	Clearly, $\partial \mathcal{M}/\partial S\neq T$; in fact, the relation is given by $\partial\mathcal{M}/\partial S=T/\Delta$, where
	\begin{equation}
		\Delta\equiv\frac{r_2^3}{(r_2^\sigma+q^\sigma)^{3/\sigma}}+\frac{b^{4n-4}q r_2^3}{2\gamma (r_2^\nu+b^\nu)^{4n/\nu}}.
	\end{equation}
	This discrepancy arises because the standard form of the first law is derived under the assumption that the Lagrangian of the theory does not explicitly depend on the $\mathcal{M}$.
	However, this assumption does not hold in the present solution.
	
	For magnetic solutions, the dependence of the NED Lagrangian in Eq.~\eqref{RMBPL} on $\mathcal{M}$ is given by
	\begin{equation}
		L=\frac{2\mathcal{M}q}{\gamma b^{4-4n} }\frac{3b^\nu+\gamma r^\nu}{(r^\nu+b^\nu)^{1+4n/\nu}}+\frac{12\mathcal{M}q^\sigma}{(r^\sigma+q^\sigma)^{1+3/\sigma}}.
	\end{equation}
	Since $m(r)=\mathcal{M}q b^{4n-4}r^3/[2\gamma (r^\nu+b^\nu)^{4n/\nu}]+\mathcal{M} r^3/(r^\sigma+q^\sigma)^{3/\sigma}$ and $L=4m^\prime(r)/r^2$, according to the method given in \cite{ma2014corrected}, integrate the later equation, we obtain
	\begin{equation}
		m(r_\infty)-m(r_2)=\frac{1}{4}\int_{r_2}^{\infty}L r^2\mathrm{d}r.
	\end{equation}
	Taking the variation of the above equation, it follows that
	\begin{equation}
		\Delta \frac{\partial \mathcal{M}}{\partial S}=\frac{1}{4\pi r_2}(1-\frac{r_2^2}{2}L)=T.
	\end{equation} 
	Apply the constraint $\mathcal{M}=\alpha^{1/4}g^{3/2}$ and treating both $b, q$ as thermodynamic variables, the first law can be expressed as
	\begin{equation}
		\Delta\mathrm{d}\mathcal{M}=T \mathrm{d}S+P_b \mathrm{d}b++P_q \mathrm{d}q,\label{65}
	\end{equation}
	where $P_b=\Delta\partial \mathcal{M}/\partial b,P_q=\Delta\partial \mathcal{M}/\partial q$.
	Clearly, for electric solutions, $\Delta$ takes the same form.
	
\begin{figure}[htbp]
		\centering
		
		\begin{subfigure}[b]{0.45\textwidth}
			\includegraphics[width=\textwidth]{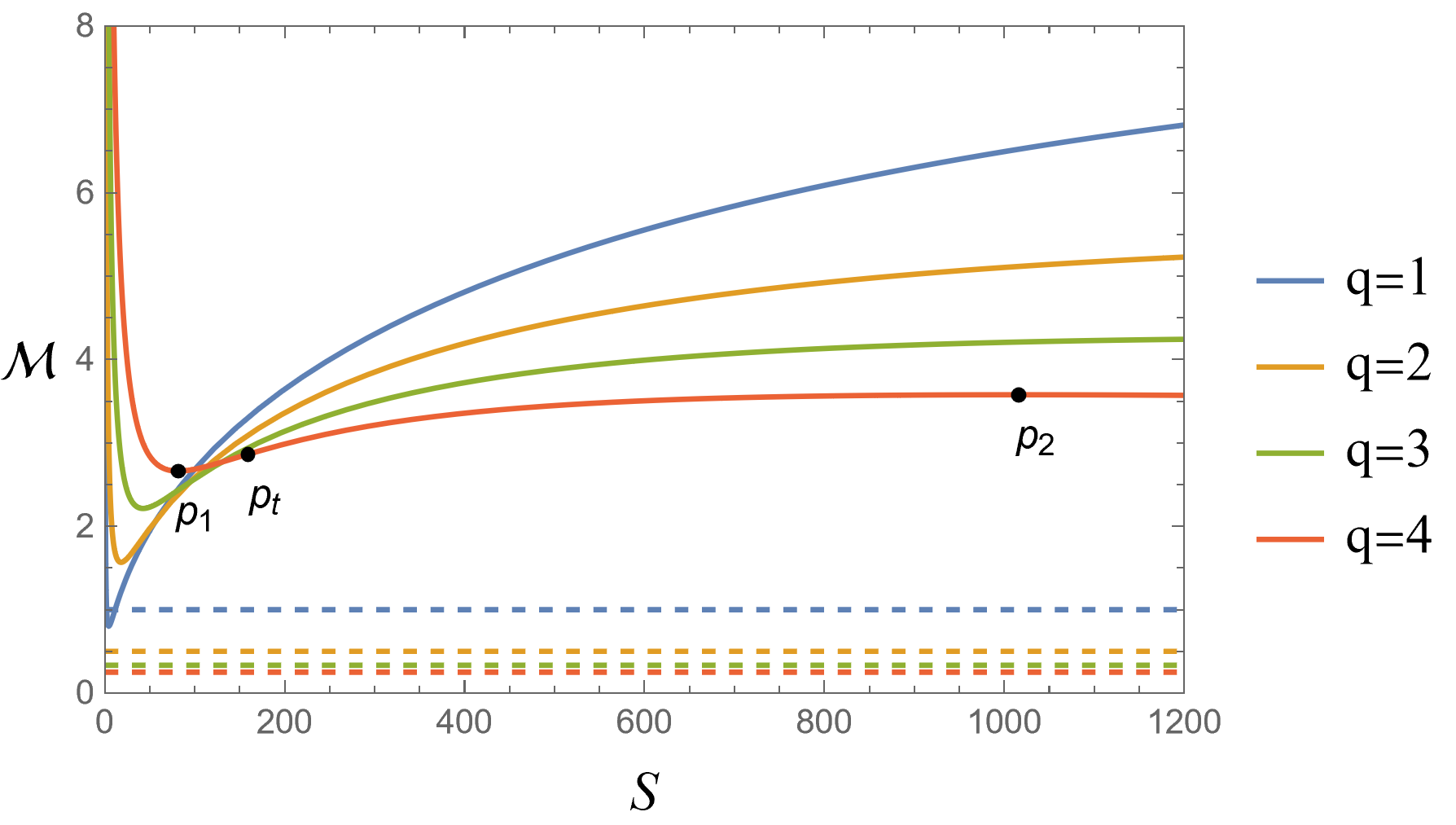}
			\caption{$n=1/2,\nu=1/4,\sigma=4,b=1$. The dashed line denotes the values of $\mathcal{M}$ as $S\to +\infty$.}
			\label{fig:y1}
		\end{subfigure}
		
		\vspace{0.2cm}
		
		\begin{subfigure}[b]{0.47\textwidth}
			\includegraphics[width=\textwidth]{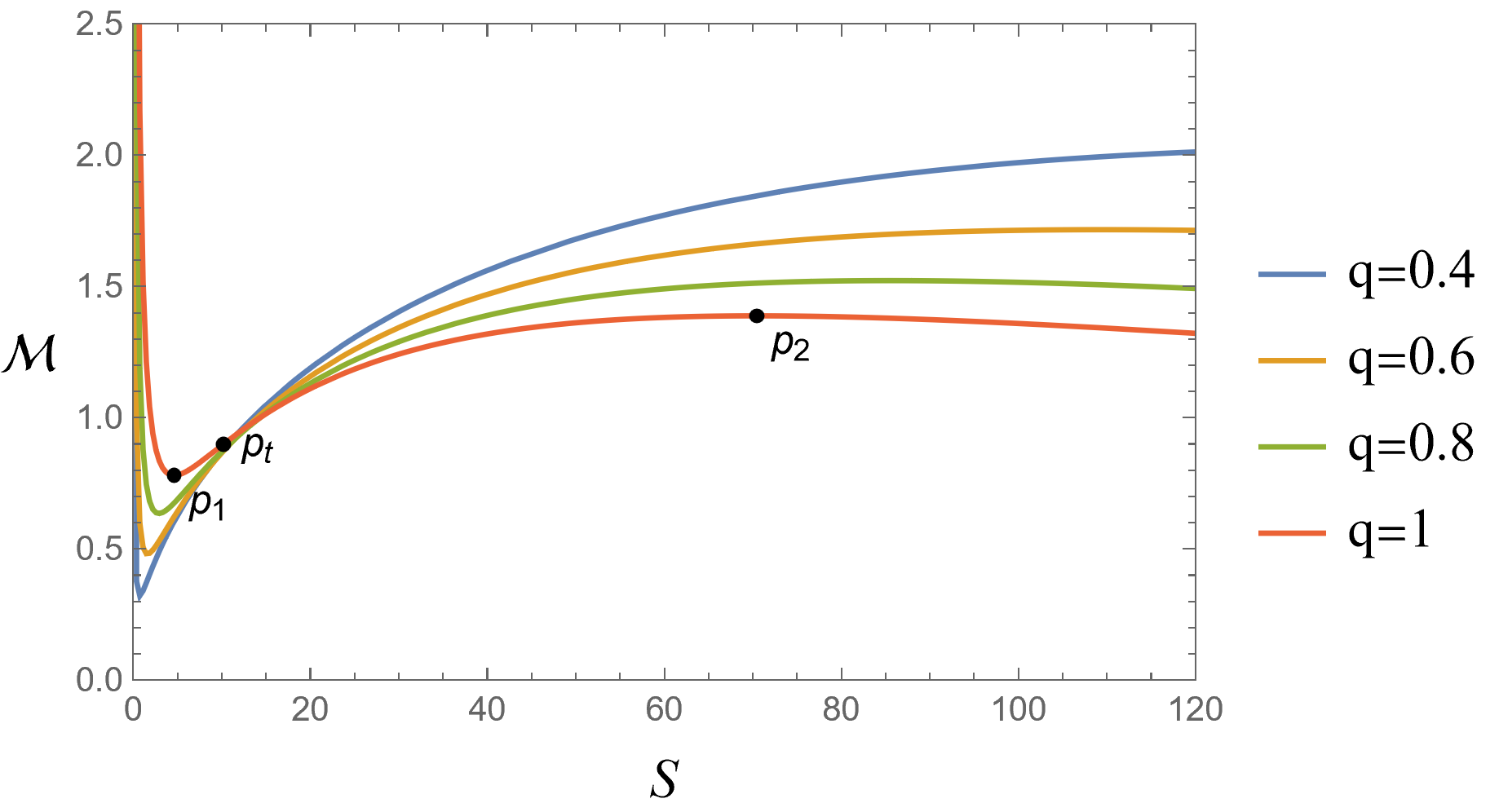}
			\caption{$n=1/4,\nu=1/4,\sigma=4,b=1$.}
			\label{fig:y4}
		\end{subfigure}
		
		\caption{The behavior of $\mathcal{M}$ associated with Eq.~\eqref{MSbq}.}
		\label{fig:y}
	\end{figure}

	The behaviour of $\mathcal{M}$ is depicted in Fig.~\ref{fig:y}.
	The two extremal points of $\mathcal{M}$ are denoted as $p_1$ and $p_2$.
	For $n=1/2$, since $\mathcal{M}\to b^2/q$ as $S\to+\infty$, black holes exist only for values of $S$ satisfying $\max(\mathcal{M}_{p_1}, b^2/q)<\mathcal{M}(S)<\mathcal{M}_{p_2}$.
	Within this range, each $m(S)$ corresponds to three distinct values of $S$, and thus three values of $r$.
	For $0<n<1/2$, since $\mathcal{M}\to 0^+$ as $S\to+\infty$, black holes exist only for values of $S$ such that $\mathcal{M}_{p_1}<\mathcal{M}(S)<\mathcal{M}_{p_2}$.
	Moreover, the temperature $T$ must be positive at $r=r_2$; hence, $T$ is physically meaningful only for $S_{p_1}<S<S_{p_2}$, due to the relationship between $T$ and $\partial \mathcal{M}/\partial S$.
	In fact, $\mathcal{M}_{p_1}$ and $\mathcal{M}_{p_2}$ correspond to those of two distinct types of extremal black holes, as previously discussed.
	Hence, there are two zeros of the heat capacity $C$ (corresponding to points where $T$ vanishes).
	Furthermore, it should be noted that the requirement $\sqrt{\lambda}g>1$ (for $n=1/2$) translates to $\mathcal{M}q b^{-2}>1$.

	\subsection{Smarr formula.}
	\label{sub:4b}
	To derive the Smarr formula, it is necessary to utilize the properties of homogeneous functions \cite{hankey1972systematic,rodrigues2022bardeen}.
	In terms of the entropy, $\mathcal{M}$ can be written as
	\begin{align}
		\mathcal{M}(S,b,q)=&\frac{\pi \gamma}{S}/\bigg\{b^{4n-4}q\left[\left(S/\pi\right)^{\nu/2}+b^\nu\right]^{-4n/\nu}\notag\\&+2\gamma\left[\left(S/\pi\right)^{\sigma/2}+q^\sigma\right]^{-3/\sigma}\bigg\}.\label{mm}
	\end{align}
	To determine the degree of homogeneity, $\mathcal{M}$ is expressed as
	\begin{align}
		\mathcal{M}(l^xS,l^yb,l^z q)=&\frac{\pi \gamma}{l^xS}/\bigg\{(l^yb)^{4n-4}(l^z q)\notag\\&\left[\left(l^xS/\pi\right)^{\nu/2}+(l^yb)^\nu\right]^{-4n/\nu}\notag\\&+2\gamma\left[\left(l^xS/\pi\right)^{\sigma/2}+(l^z q)^\sigma\right]^{-3/\sigma}\bigg\}.
	\end{align}
	To isolate $l$, the assumptions $y=z=x/2$ and $x=1$ are made, yielding
	\begin{align}
		\mathcal{M}(lS,\sqrt{l}b,\sqrt{l}q)=&\frac{\pi \gamma\sqrt{l}}{S}/\bigg\{b^{4n-4}q\left[\left(S/\pi\right)^{\nu/2}+b^\nu\right]^{-4n/\nu}\notag\\&+2\gamma\left[\left(S/\pi\right)^{\sigma/2}+q^\sigma\right]^{-3/\sigma}\bigg\},
	\end{align}
	such that $\mathcal{M}$ is a homogeneous function with degree
	of homogeneity $1/2$.
	According to Euler’s identity \cite{hankey1972systematic,rodrigues2022bardeen}, it follows that
	\begin{equation}
		\frac{1}{2}\mathcal{M}=\frac{\partial \mathcal{M}}{\partial S} S+\frac{1}{2}\frac{\partial \mathcal{M}}{\partial b}b+\frac{1}{2}\frac{\partial \mathcal{M}}{\partial q}q.
	\end{equation}
	Therefore, the Smarr formula can be wrriten as
	\begin{equation}
		\Delta\mathcal{M}=2TS+P_b b+P_q q.\label{71}
	\end{equation}
	It can be seen that this result is consistent with the expression given in Eq.~\eqref{mm}.
	
				\begin{figure*}[htbp]
		\centering
		
		\begin{subfigure}[b]{0.45\textwidth}
			\includegraphics[width=\textwidth]{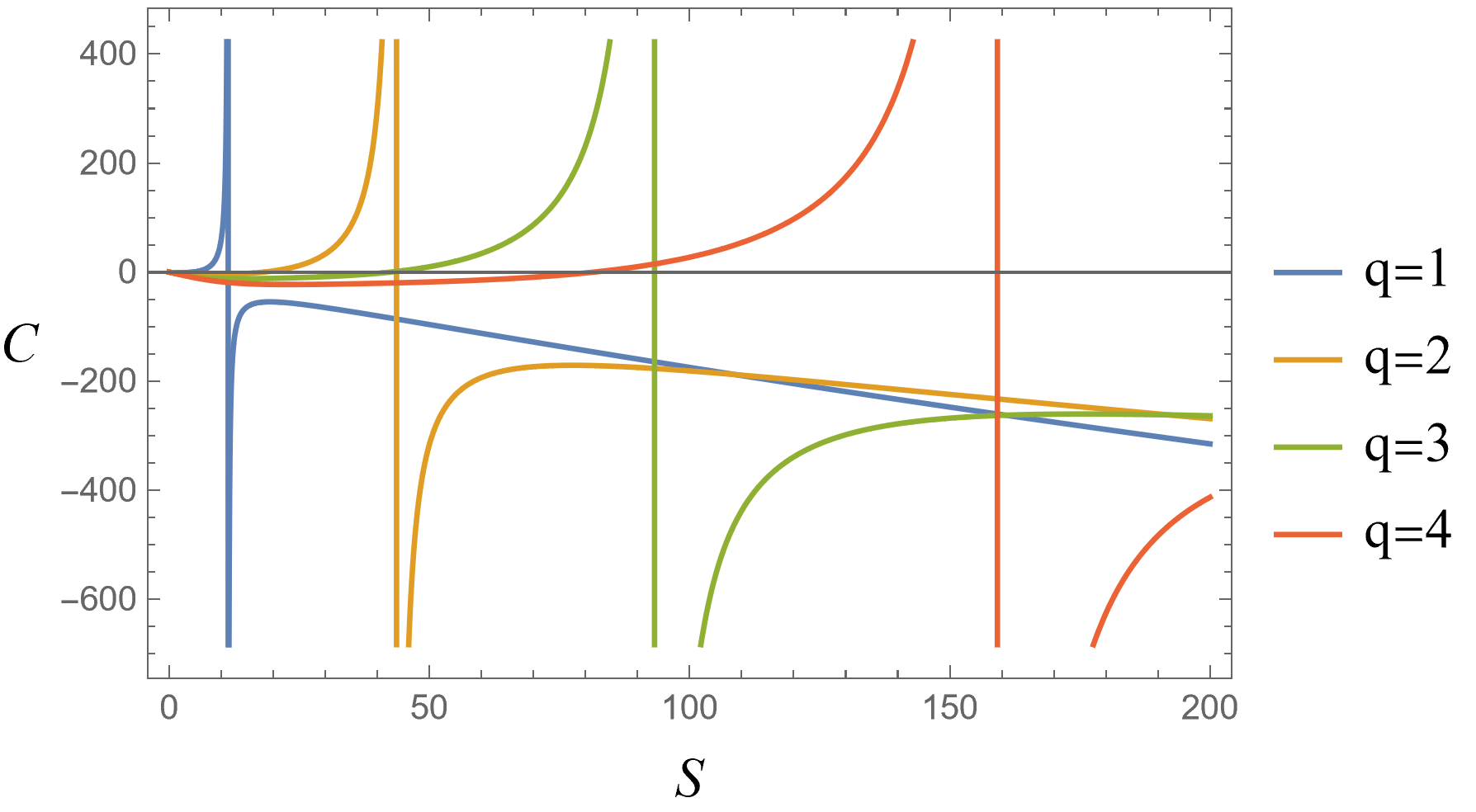}
			\caption{$n=1/2,\nu=1/4,\sigma=4,b=1$.}
			\label{fig:y2}
		\end{subfigure}
		\hfill
		\begin{subfigure}[b]{0.45\textwidth}
			\includegraphics[width=\textwidth]{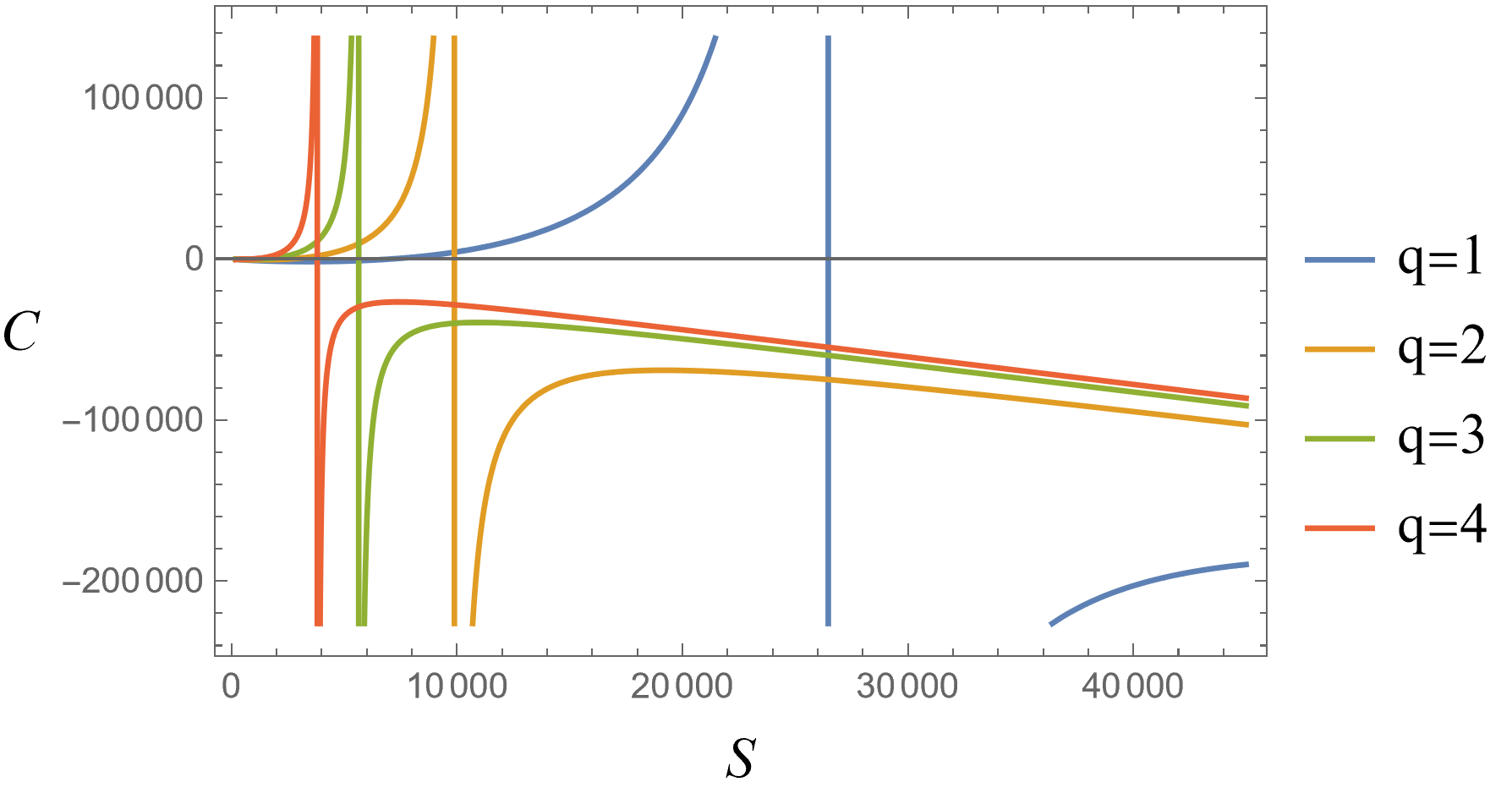}
			\caption{$n=1/2,\nu=1/4,\sigma=4,b=1$.}
			\label{fig:y3}
		\end{subfigure}
		
		\vspace{0.2cm}
		
		\begin{subfigure}[b]{0.45\textwidth}
			\includegraphics[width=\textwidth]{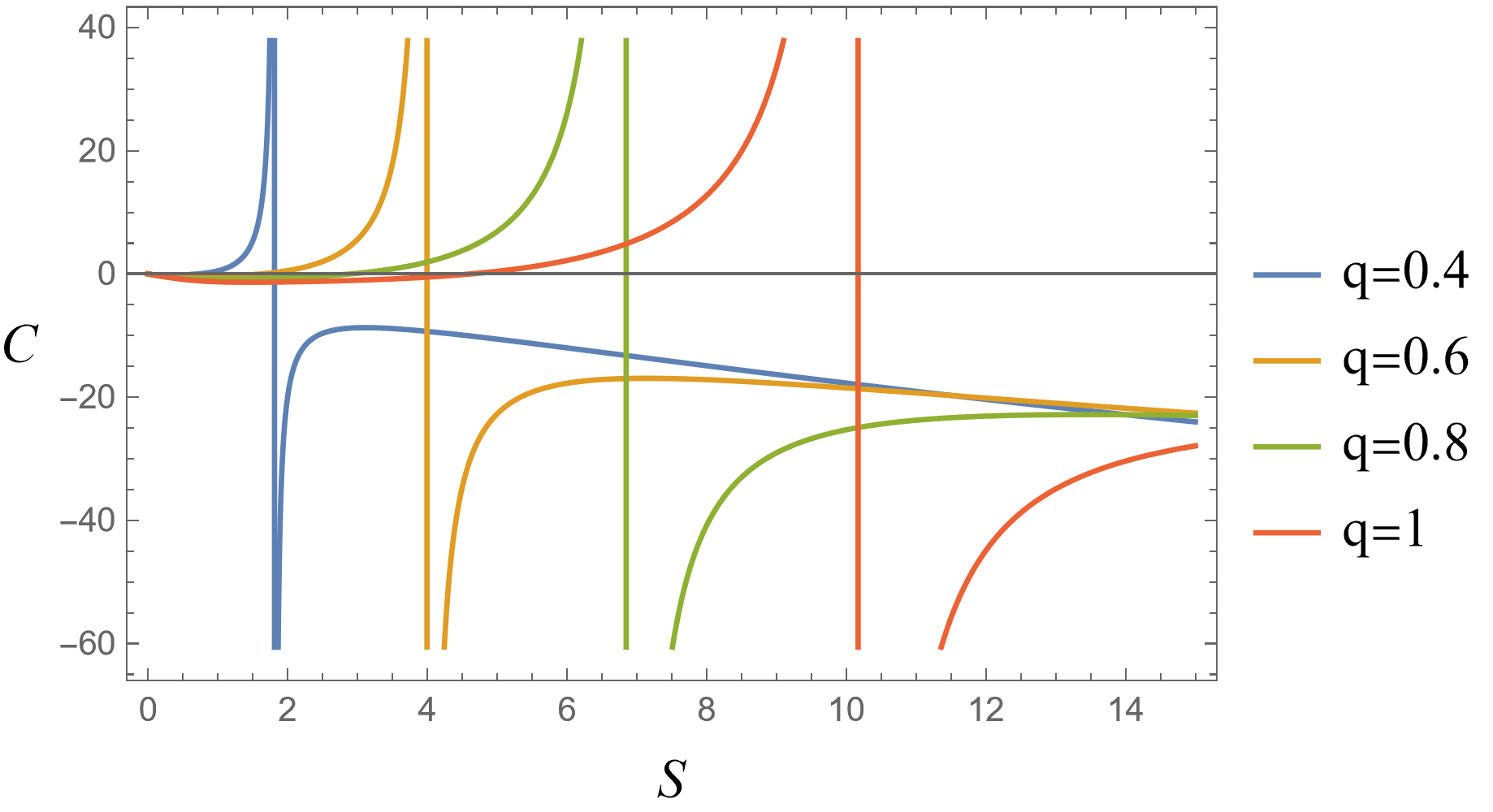}
			\caption{$n=1/4,\nu=1/4,\sigma=4,b=1$.}
			\label{fig:y5}
		\end{subfigure}
		\hfill
		\begin{subfigure}[b]{0.45\textwidth}
			\includegraphics[width=\textwidth]{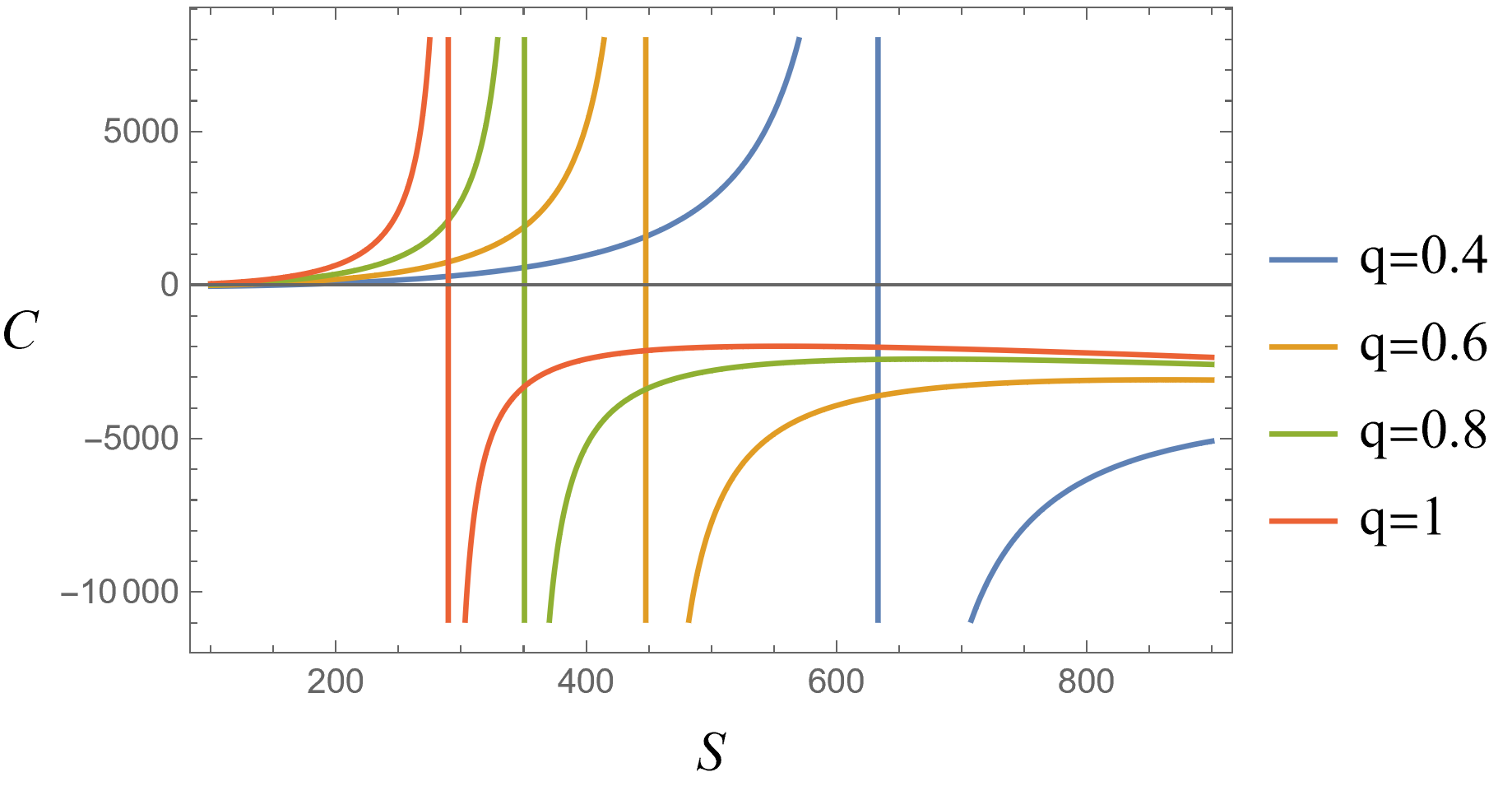}
			\caption{$n=1/4,\nu=1/4,\sigma=4,b=1$.}
			\label{fig:y6}
		\end{subfigure}
		
		\caption{The behavior of the heat capacity $C$.}
		\label{fig:y2356}
	\end{figure*}

	\subsection{Heat Capacity.}
	\label{sub:4c}
	To analyze the local stability of the solutions, the black hole heat capacity is considered, which is given by
	\begin{equation}
		C=T\left(\frac{\partial S}{\partial T}\right)_{b,q}.
	\end{equation}
	Using the expressions for entropy and temperature presented above, the heat capacity can be expressed as a function of $S$; however, due to its length, the explicit form is omitted here.
	
	The behaviour of heat capacity $C$ is shown in Fig.~\ref{fig:y2356}.
	It can be observed that, whether $n=1/2$ or $n=1/4$, two phase transitions occur.
	The entropy at the first phase transition is denoted by $S_{p_t}$.
	Therefore, as shown in Fig.~\ref{fig:y}, for $S_{p_1}<S<S_{p_t}$, a black hole exists with positive heat capacity, corresponding to a stable black hole solution.
	
	\section{Conclusion.}
	\label{con}
	In this study, we present a new class of regular spacetimes that are asymptotically dynamical as described in Eq.~\eqref{RPMB}, yet not asymptotically de Sitter, and subsequently extend these solutions to regular black holes.
	It is demonstrated that these solutions originate from nonlinear electrodynamics theories asymptotically characterized by power-Maxwell behavior as in Eq.~\eqref{PL}.
	Notably, the solutions belong to distinct nonlinear theories in the magnetic and electric cases.
	The magnetic solutions given in Eq.~\eqref{RMBPL} are derived, and their causal structures are analyzed through Penrose diagrams, illustrating the transition from power-Maxwell solutions (Figs.~\ref{fig:1} and~\ref{fig:2}) to regular power-Maxwell solutions (Fig.~\ref{fig:2}, featuring a topology change), and ultimately to regular power-Maxwell black holes (Fig.~\ref{fig:6}).
	Various parameter regimes admitting black holes are illustrated in Figs.~\ref{fig:z} and~\ref{fig:l}.
	By applying the well-known FP-duality transformation in Eq.~\eqref{FP}, the corresponding electric solutions are obtained within the $P$ framework.
	Uniqueness conditions for the electric solutions within the $F$ framework are identified, specifically $0<\nu \leq \gamma$ when $m=0$.
	For cases $m \neq 0$, and the presence of a black hole, uniqueness appears to fail, as demonstrated in Fig.~\ref{fig:c}.
	Interestingly, a magnetic solution exists in the square-root Maxwell theory (for $n=1/2$), but no corresponding dual electric solution is found.
	To address this, we reformulate the nonlinear theory in terms of an auxiliary field as introduced in Eqs.~\eqref{LFP} and~\eqref{PS}, and extend the theory to accommodate such solutions Eq.~\eqref{TFP}.
	The duality is then derived in this auxiliary field representation Eq.~\eqref{PDU}.
	Next, the effective metrics corresponding to the magnetic solution in Eq.~\eqref{51} and the electric solution in Eq.~\eqref{59} are analyzed.
	For magnetic configurations, if $m=0$, the effective metric remains free of singularities under the condition $0<\nu \leq \gamma$.
	In cases where $m \neq 0$, and a black hole is present, singularities in the effective metric seems like to appear, as illustrated in Fig.~\ref{fig:c}.
	For electric solutions, singularities in the effective metric persist even when $m=0$.
	When black holes exist, the effective metric may contain multiple singularities.
	Furthermore, for both magnetic and electric solutions, spacelike photon trajectories exist.
	Finally, it is demonstrated that the ADM mass enters the Lagrangian, and the first law of black hole thermodynamics (Eq.\eqref{65}) along with the associated Smarr formula (Eq.\eqref{71}) are derived for the system.
	The thermodynamic stability is examined through the behavior of the heat capacity (Figs.\ref{fig:y} and~\ref{fig:y2356}), demonstrating that stable black hole solutions with positive heat capacity indeed exist.
	
	The existence of spacelike photons and the associated causal structure issues warrant further in-depth investigation.
	Additionally, other regular black holes, such as black-bounce spacetimes, deserve systematic study \cite{simpson2019black,lobo2021novel,franzin2021charged,bronnikov2022black,bronnikov2022field,canate2022black}.
	Studies on early cosmic acceleration potentially driven by regular black holes have contributed to increased attention towards the investigation of regular black holes \cite{dialektopoulos2025primordial}.
	
	\begin{acknowledgments}
		The authors gratefully acknowledge Professor Qi-Yuan Pan for valuable discussions during the preparation of this manuscript.
	\end{acknowledgments}
	
\bibliography{c}

\end{document}